\documentclass{aa}
\usepackage{graphicx}
\usepackage{subcaption}
\usepackage{enumitem}
\usepackage[dvipsnames]{xcolor}
\usepackage{amsmath}
\usepackage{txfonts}
\usepackage{soul}
\usepackage[colorlinks=True,pdfborder={0 0 0}, linkcolor=blue,citecolor=blue]{hyperref}

\begin{document}

   \title{Numerical Insights into Disk Accretion, Eccentricity, and Kinematics in the Class 0 phase}
   \author{Adnan Ali Ahmad\inst{1}
          \and
          Benoît Commerçon\inst{1}
          \and
          Elliot Lynch\inst{1}
          \and
          Francesco Lovascio\inst{1}
          \and
          Sebastien Charnoz\inst{2}
          \and
          Raphael Marschall\inst{3}
          \and
          Alessandro Morbidelli\inst{4,5}
          }

   \institute{ENS de Lyon, CRAL UMR5574, Universite Claude Bernard Lyon 1, CNRS, Lyon 69007, France
   \and Université Paris Cité, Institut de Physique du Globe de Paris, CNRS, Paris, F-75005, France
   \and International Space Science Institute, Hallerstrasse 6 3012 Bern, Switzerland
   \and Collège de France, CNRS, PSL Univ., Sorbonne Univ., Paris, 75014, France
   \and Observatoire de la Côte d’Azur, Université Cote d’Azur, CNRS, Laboratoire Lagrange, 06304 Cedex 4 Nice, France
    }
   \date{Received XXXX; accepted XXXX}

 
  \abstract
  {The formation and early evolution of protoplanetary disks are governed by a wide variety of physical processes during a gravitational collapse. Observations have begun probing disks in their earliest stages, and have favored the magnetically-regulated disk formation scenario. Disks are also expected to exhibit ellipsoidal morphologies in the early phases, an aspect that has been widely overlooked.}
   {We aim to describe the birth and evolution of the disk while accounting for the eccentric motions of fluid parcels.}
   {Using 3D radiative magnetohydrodynamic (MHD) simulations with ambipolar diffusion, we self-consistently model the collapse of isolated $1~\mathrm{M_\odot}$ and $3~\mathrm{M_\odot}$ cores to form a central protostar surrounded by a disk. We account for dust dynamics, and employ gas tracer particles to follow the thermodynamical history of fluid parcels.}
   {We find that magnetic fields and turbulence drive highly anisotropic accretion onto the disk via dense streamers. This streamer-fed accretion, occurring from the vertical and radial directions, drives vigorous internal turbulence that facilitates efficient angular momentum transport and rapid radial spreading. Crucially, the anisotropic inflow delivers material with an angular momentum deficit that continuously generates and sustains significant disk eccentricity ($e\sim 0.1$).}
   {Our results reveal ubiquitous eccentric kinematics in Class 0 disks, with direct implications for disk evolution, planetesimal formation, and the interpretation of cosmochemical signatures in Solar System meteorites.}


   \maketitle
%

\section{Introduction}

Recent advances in FIR/sub-mm observations (ALMA, NOEMA, VLA) and MHD simulations have improved our understanding of protoplanetary disks. However, key gaps remain, particularly in the Class 0 phase where disks, which are deeply embedded in their nascent cores, are challenging to observe and simulate due to highly non-linear processes acting across a large dynamical range (see the reviews by \citealp{teyssier_2019, tsukamoto_2023b, lesur_2023}). The magnetically-regulated disk-formation scenario \citep{hennebelle_2016, wurster_2018, vaytet_2018, wurster_disks, wurster_2020, lee_2021, lebreuilly_2023b, mayer_2024, lee_2024} is the favored theoretical model, where the envelope's magnetic field extracts angular momentum and ambipolar diffusion regulates magnetic braking. This produces disk radii of $<\sim50$ AU, consistent with the observations of \cite{maury_2019, tobin_2020}\footnote{Mass estimates remain uncertain, see \cite{bergin_2017, duy_2024, savvidou_2025}.}. Said models predict $\sim0.1\ \mathrm{G}$ field strengths to reproduce observations, which aligns with paleomagnetic measurements of solar system meteorites \citep{weiss_2021}. 
Once formed, Class 0 disks experience vigorous accretion in the form of infall that shapes their dynamical evolution. Understanding this phase is crucial for planet formation scenarios, as current surveys find Class II disk masses insufficient to explain exoplanetary systems \citep{greaves_2010, williams_2012, najita_2014, mulders_2015, mulders_2018}. This implies either early planet formation during the Class 0 phase, or continuous mass replenishment from the surroundings \citep{manara_2018, padoan_2025}. In addition, cosmochemical analyses of meteorites have provided constraints on disk kinematics in the early solar system \citep{morbidelli_2024}, indicating that calcium aluminum inclusions (CAIs) formed near the Sun and were transported to larger distances where they were incorporated into carbonaceous chondrites. Modeling this early phase by describing accretion processes and their kinematic impacts on the disk is therefore essential to bridge the gap between star and planet formation models.
\\
Several studies have investigated disk formation during gravitational collapse. \cite{lee_2021} simulated the collapse of a $1\ \mathrm{M_{\odot}}$ cloud using radiative-MHD (including flux limited diffusion and ambipolar diffusion), and found that magnetic fields dominate infall dynamics and that the disk midplane contributes less to mass transport than the upper layers. Similarly, \cite{tu_2024} studied a magnetized collapse of a $0.5\ \mathrm{M_{\odot}}$ cloud core with ambipolar diffusion, albeit by adopting a barotropic equation of state. They found that gravo-magnetic sheets feed the disk primarily through its upper and lower surfaces. \cite{xu_2021b} similarly carried out a collapse with ambipolar and Ohmic diffusion, albeit in a non-turbulent core, and found that angular momentum transport within the disk is driven by gravitational instabilities. On larger scales, \cite{mayer_2025} and \cite{yang_2025} performed zoom-in barotropic MHD simulations from cloud-scale initial conditions. While \cite{yang_2025} used ideal MHD, \cite{mayer_2025} included ambipolar diffusion. Both found that magnetic fields and inherited cloud-scale turbulence channel accretion into specific regions in the disk. On the high-mass end, \cite{he_2023, he_2025} carried out similar zoom-in simulations including ideal MHD and radiative transfer, highlighting how turbulence mitigates magnetic braking. Across both low and high-mass regimes, anisotropic accretion has been consistently reported, and has been further connected to the development of eccentric motions within disks \citep{commercon_2024, calcino_2025}. In addition, external accretion can drive turbulence in the disk, thereby transporting angular momentum \citep{lesur_2015, hennebelle_2016b, hennebelle_2017, Kuznetsova_2022}.
\\
In this study, we present a comprehensive analysis of the accretion and kinematics of protoplanetary disks formed through self-consistent radiative-MHD simulations of gravitational collapse in an isolated dense core. We characterize how anisotropic accretion shapes both the global disk kinematics and the previously unexplored dynamics of eccentric fluid motions, an important yet overlooked aspect of early disk evolution.

\section{Model} \label{section:model}

We model the collapse of an isolated uniform-density core of mass $M_0$ using the {\ttfamily RAMSES} adaptive mesh refinement (AMR) code \citep{teyssier_2002}. The core is initialized with a temperature of $T_0 = 10~\mathrm{K}$ and size $R_{0}$ (thermal-to-gravitational energy ratio $\alpha = 0.4$), solid-body rotation along the $-z$ axis (rotational-to-gravitational energy ratio $\beta = 4 \times 10^{-2}$), and a uniform magnetic field tilted by $10^\circ$ relative to the rotation axis, with a mass-to-flux ratio $\mu = (10/3)\mu_{\mathrm{crit}}$. No initial turbulent velocity field was present in this run; instead, we introduce an m=2 azimuthal density perturbation of amplitude 10\%. In this setup, the magnetic interchange instability develops spontaneously during the collapse.\footnote{This instability generates a turbulent-like velocity field which, while different from models that include an inherited Kolmogorov cascade from larger scales (see \citealp{kuffmeier_2024}), leads to significant anisotropic accretion.}
\\
The collapse proceeds self-consistently until the density exceeds $5 \times 10^{12}~\mathrm{cm^{-3}}$, at which point a sink particle forms \citep{bleuler_2014}, accreting material within a radius of $4\Delta x_\mathrm{min}$ (where $\Delta x_\mathrm{min}$ is the resolution at the finest AMR level). The sink particle then accretes 10\% of the material whose density exceeds $5 \times 10^{12}~\mathrm{cm^{-3}}$ (see \citealp{hennebelle_disks}). The simulation begins with a $64^3$ coarse grid ($\ell_\mathrm{min} = 6$) and is refined to maintain 20 cells per Jeans length, computed with a temperature cap of 100 K. The simulation domain is $\approx 16000\ \mathrm{AU^3}$ for R1 and $\approx 48000\ \mathrm{AU^3}$ for R2.
\\
Our radiative-MHD implementation uses a hybrid solver \citep{mignon_2021b,mignon_2021a}, combining the M1 method for stellar irradiation from sink particles \citep{Rosdahl_2013,Rosdahl_2015} with gray flux-limited diffusion for absorption and radiative transfer \citep{commercon_2011,commercon_2014}. Non-ideal MHD effects are present in the form of ambipolar diffusion \citep{Fromang_2006, teyssier_2006, masson_2012}.
\\
Two runs are presented in this paper: R1 and R2. R1 has $M_{0}=1\ \mathrm{M_{\odot}}$ and $\ell_{\mathrm{max}}=14$ ($\Delta x_{\mathrm{min}}\approx 0.97~\mathrm{AU}$). R2 has $M_{0}=3\ \mathrm{M_{\odot}}$ and $\ell_{\mathrm{max}}=16$ ($\Delta x_{\mathrm{min}}\approx 0.72~\mathrm{AU}$). Assuming a $\approx 30\%$ star formation efficiency \citep{andre_2010, konyves_2015}, these should form a 0.3 $\mathrm{M_{\odot}}$ and 1 $\mathrm{M_{\odot}}$ star, respectively. R1 was previously presented in \cite{commercon_2024} and is similar to \cite{hennebelle_disks, lee_2021}. The spatial resolution of both runs is traded for extended integration time, enabling us to study the long-term evolution of the disk.
\\
We also account for dust dynamics under the monofluid approximation \citep{lebreuilly_2019, commercon_2023}. We consider $\mathcal{N}_{\mathrm{d}}$ bins of single-size dust species with sizes ranging from $10^{-3}~\mathrm{\mu m}$ to $20~\mathrm{\mu m}$. Dust particles do not undergo coagulation or fragmentation. We use $\mathcal{N}_{\mathrm{d}}=10$ for R1 and $\mathcal{N}_{\mathrm{d}}=5$ for R2.
\\
Finally, $10^{4}$ massless Lagrangian particles are used in run R2 to trace the gas dynamics during the collapse using the particle-mesh method, allowing us to track the temperature-density history of gas parcels during the collapse. These are randomly distributed throughout the volume of the initial dense core. Run R1 does not contain any tracer particles.

\section{Global disk \& ejecta properties}

\begin{figure*}[h]
\centering
\includegraphics[scale=.36]{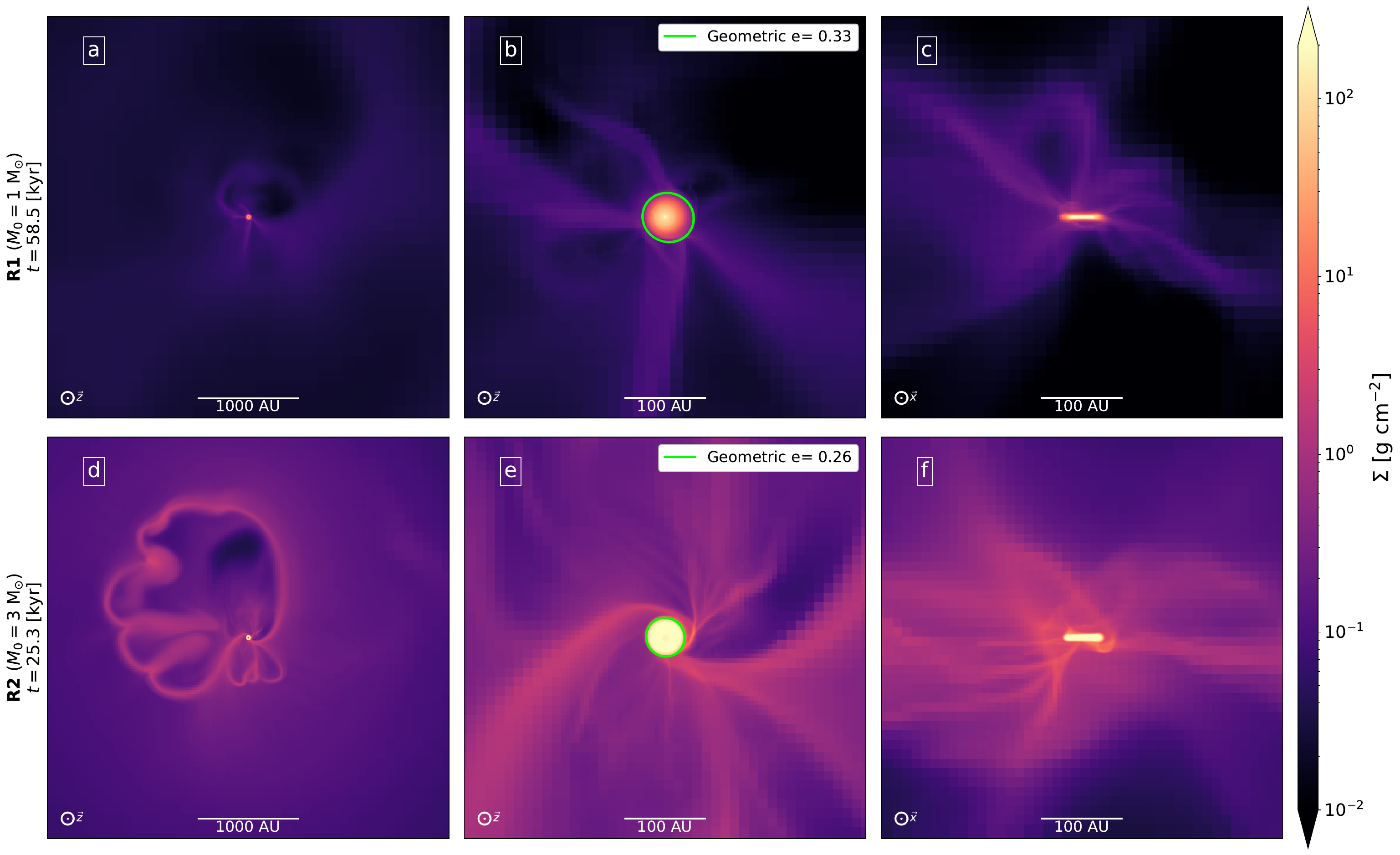} 
\caption{Column density images in the $z$ (a, b, d, e) and $x$ (c, f) directions for runs R1 (first row) and R2 (second row), at both large scales (a, d) and small scales (b, c, e, f). The lime colored contours in panels (b,e) are a least-square fit of an ellipse enveloping the disk's surface. The associated eccentricity is shown in the top-right corner of each panels (b,e).}
\label{fig:coldens}
\end{figure*}

We begin by presenting the qualitative and quantitative properties of the disks formed in runs R1 and R2. Figure~\ref{fig:coldens} shows column density images at $t=58.5$~kyr for R1 and $t\approx27$~kyr for R2, with $t=0$ corresponding to the epoch of sink formation. In both runs, strong radial magnetic field gradients trigger the interchange instability \citep{spruit_1995}, visible at large scales (Fig.~\ref{fig:coldens}a,d) as distinct streamers feeding the disks. This highlights the magnetic field's dominant role in governing gas kinematics during the collapse and subsequent disk accretion. The instability is noticeably more pronounced in R2 due to its greater mass reservoir. Edge-on views (Fig.~\ref{fig:coldens}c,f) reveal numerous streamers connecting to the disk, demonstrating anisotropic accretion. As discussed in Section~\ref{sec:e}, this anisotropic accretion directly drives eccentric motions within the disk (see also \citealp{commercon_2024, calcino_2025}).
\\
Figure~\ref{fig:coldens}b,e shows an ellipse fitted to the disk's outer edge\footnote{Done using OpenCV's \citep{opencv_library} implementation of \cite{fitzgibbon_1995}'s algorithm.}, from which we measure the geometric eccentricity via $\sqrt{1-\frac{b^2}{a^2}}$ ($a$ and $b$ being the major/minor axes). The disk exhibits significant eccentricity, with values of 0.33 for R1 and 0.26 for R2, with streamers feeding its outer regions anisotropically\footnote{These are different from the late Class 2 streamers discussed in the literature (e.g., \citealp{pineda_2020, Cacciapuoti_2024}), which originate from larger scales than that of the dense core.}. Figure~\ref{fig:globalprops} tracks the temporal evolution post-sink formation (which marks our $t=0$)\footnote{See Appendix \ref{appendix:diskdef} for the disk definition.}. Both runs show qualitatively similar protostellar mass growth ($M_{*}$, Fig.~\ref{fig:globalprops}a), though R2's rate is higher. The disk mass ($M_{\mathrm{d}}$, Fig.~\ref{fig:globalprops}b) differs markedly: R1 declines steadily from $\approx\!2\!\times\!10^{-1}\,\mathrm{M_\odot}$ to $\approx\!3\!\times\!10^{-2}\,\mathrm{M_\odot}$ over 60~kyr, while R2 shows large fluctuations driven by the interchange instability. The disk radius $R_{\mathrm{d}}$ (90\% mass radius as done in \citealp{mignon_2021a}, solid lines in Fig.~\ref{fig:globalprops}b) generally follows $M_{\mathrm{d}}$'s behavior. For R1, $R_{\mathrm{d}}$ increases despite mass loss, which is consistent with viscous disk theory where outward angular momentum transport causes inward mass transport\footnote{Analyzed in Sec.~\ref{sec:masstransport}.}. The semi-major axis $a_{\mathrm{d}}$ obtained from elliptical fitting (Fig.~\ref{fig:globalprops}b) mirrors the trends in $R_{\mathrm{d}}$ but is systematically larger, as it traces the outermost orbits. The geometric eccentricity (Fig.~\ref{fig:globalprops}c) fluctuates markedly (shaded regions) due to dynamical instabilities (spiral waves, accretion streamers) altering the disk's morphology. However, the time-averaged eccentricity (solid lines) remains stable ($>0.2$) with smaller variations in R1 than R2. 
\\
Overall, R2 seems to display more dynamical behavior over time than R1. We link R2's large dip in $R_{\mathrm{d}}$ at $t\approx 15$ kyr with an influx of material with opposite angular momentum to that of the disk in Appendix \ref{appendix:shrink}.

\begin{figure}[h]
\centering
\includegraphics[scale=.34]{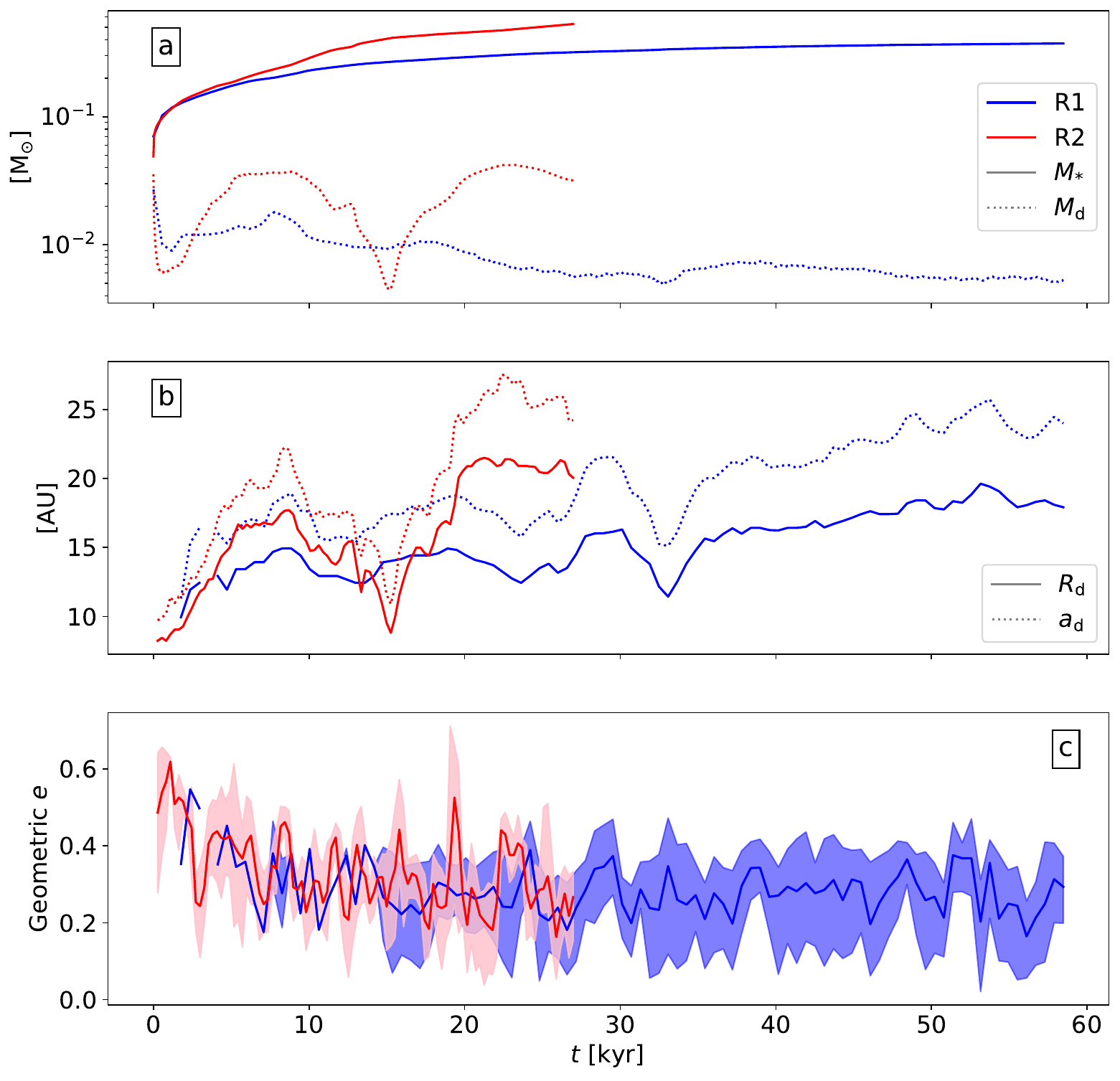} 
\caption{Global evolution of the protoplanetary disk as a function of time, where $t=0$ represents the epoch of sink (i.e., protostellar) formation. The quantities shown are the protostellar mass (solid lines in panel a), disk mass (dotted lines in panel a), disk radius (solid lines in panel b), disk semi-major axis (dotted lines in panel b), and apparent disk eccentricity (c). The disk semi-major axis and apparent eccentricity are inferred from an elliptical fit. The blue (resp. red) curve corresponds to the 1 $\mathrm{M_{\odot}}$ (resp. 3 $\mathrm{M_{\odot}}$) run R1 (resp. R2). The shaded regions in panel (d) represent temporal fluctuations in the measurement, and the solid line is an average value.}
\label{fig:globalprops}
\end{figure}

\section{Disk accretion}

The apparent eccentricities, morphological variability, and anisotropic accretion revealed in Figs.~\ref{fig:coldens} and \ref{fig:globalprops} naturally raise the question of how material is actually delivered to and redistributed within the disk. In this section, we examine the accretion process in detail by quantifying both its directional dependence and the relative contributions of vertical and radial infall. All analyses here use a reference frame centered on the sink particle, with the disk’s angular momentum defining the $z$-axis.

\subsection{Directional mass fluxes}
\label{sec:dirmf}

\begin{figure*}[h]
\centering
\includegraphics[scale=.25]{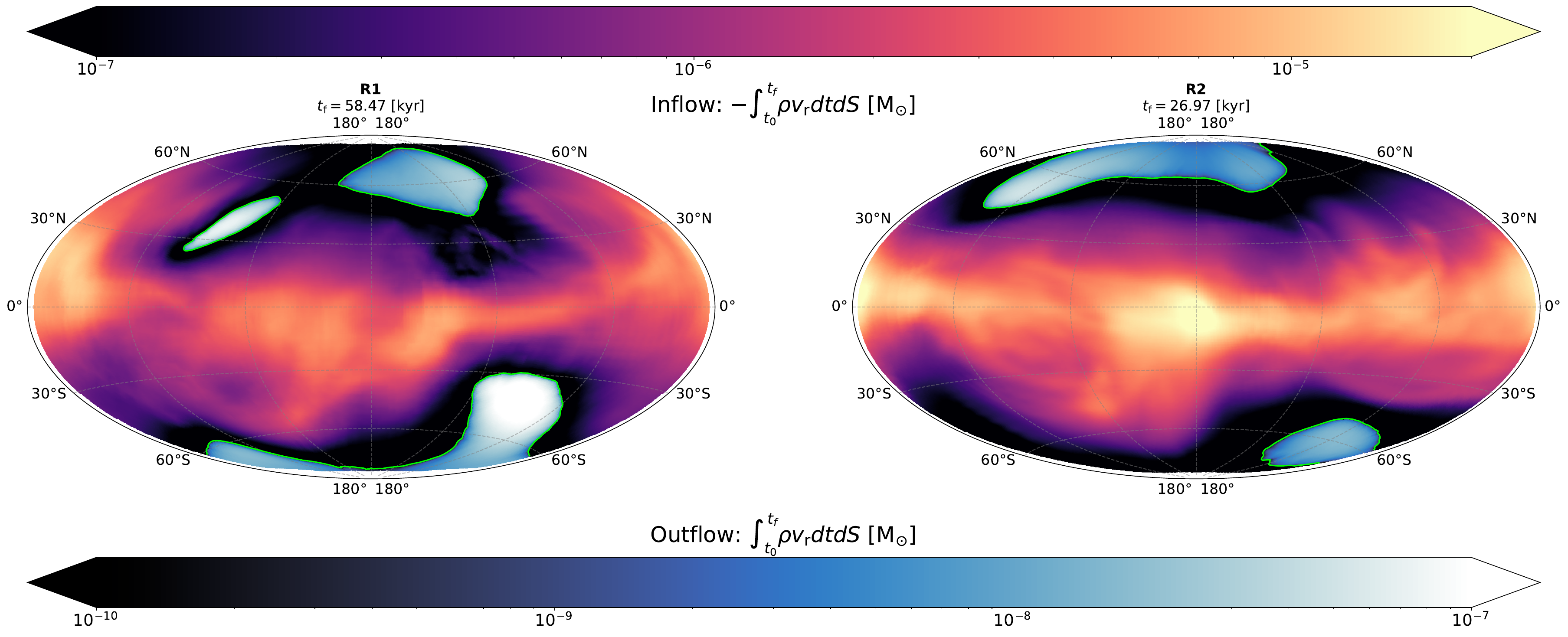} 
\caption{Hammer projections displaying the time and surface integrated radial mass flux on a sphere of $R_{\mathrm{shell}}=~30$~AU for runs R1 (left) and run R2 (right), displaying both the inflow (top colorbar) and outflow (bottom colorbar) of material throughout the simulation's duration post sink formation ($t_{\mathrm{f}}$), computed by integrating Eq. \ref{eq:mdotsphere} in time. The lime colored contour delimitates the transition from positive to negative values.}
\label{fig:shellacc}
\end{figure*}

\begin{figure}[h]
\centering
\includegraphics[scale=.42]{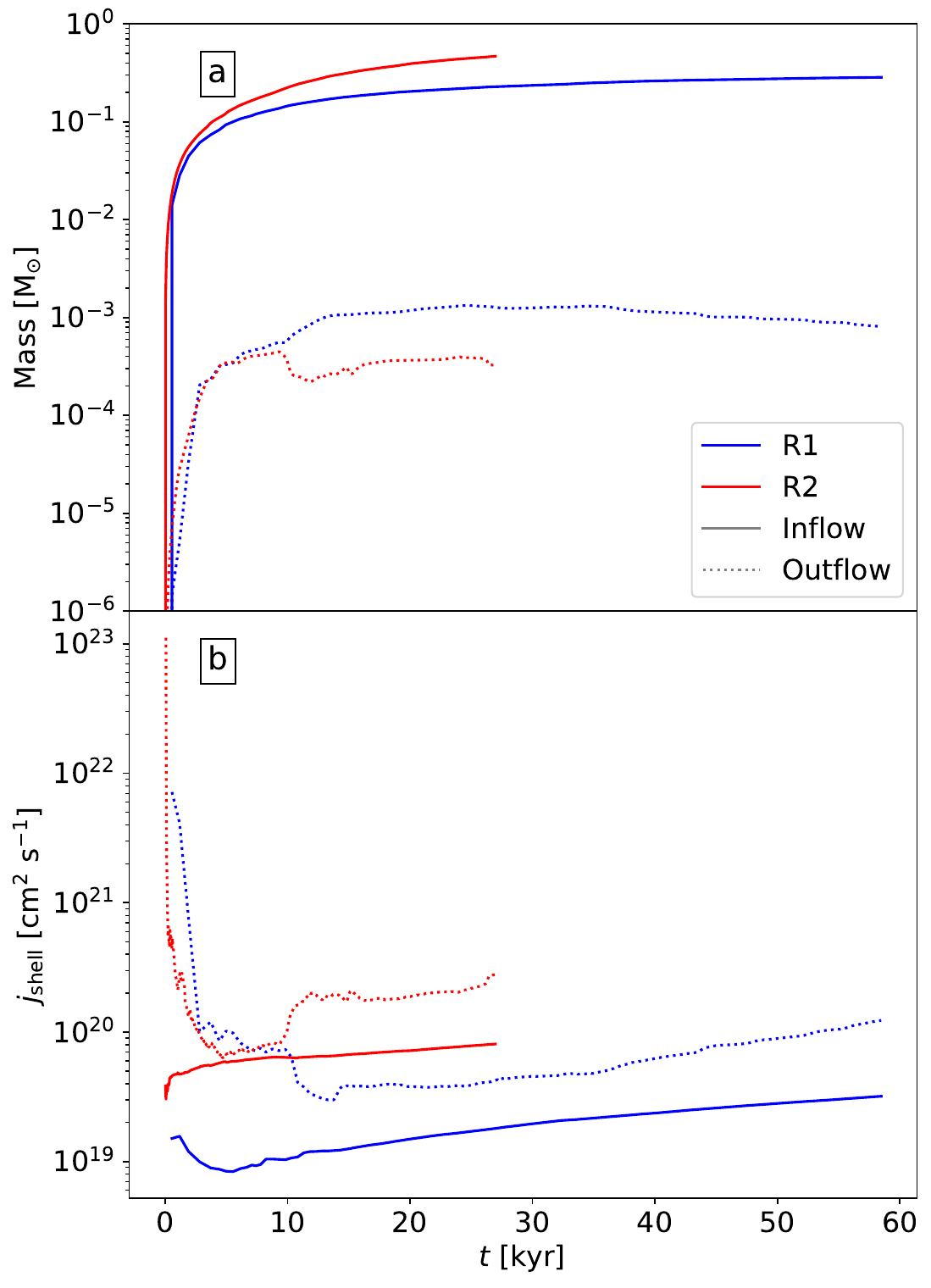} 
\caption{Total inflow (solid lines) and outflow (dotted lines) of mass (panel a) and specific angular momentum (panel b) through a sphere of radius $R_{\mathrm{shell}}=~30$~AU around the sink particle as a function of time, where $t=0$ corresponds to the epoch of sink formation, for run R1 (blue) and R2 (red). This figure is complementary to Fig. \ref{fig:shellacc}.}
\label{fig:inandout}
\end{figure}

We quantify the mass carried by the magnetically-driven streamers shown in Fig.~\ref{fig:coldens} by analyzing mass fluxes across a fixed $R_{\mathrm{shell}}=30$~AU sphere (exceeding both runs' maximum semi-major axes) around the sink particle. This approach avoids the complexities of tracking the disk's irregular surface (Fig.~\ref{fig:globalprops}c). We compute mass fluxes through spherical surface elements via
\begin{equation}
    \label{eq:mdotsphere}
    \dot{M}(\Phi, \Lambda, t) = -\rho v_{\mathrm{r}} dS = -\rho v_{\mathrm{r}} R_{\mathrm{shell}}^{2} \cos(\Phi)d\Phi d\Lambda,
\end{equation}
where $\Phi$ and $\Lambda$ are the latitude and longitude, $\rho$ is the gas density, and $v_r$ is the radial velocity in spherical coordinates.
\\
The time-integrated Hammer projections of Fig.~\ref{fig:shellacc} reveal anisotropic accretion patterns, with concentrated midplane lobes corresponding to streamer-disk connections. The polar regions show low-velocity outflows (negative fluxes) of magnetic origin, carrying $\sim100$ times less mass than the inflows (positive fluxes) (Fig.~\ref{fig:inandout}a), predominantly within the first $\sim$5~kyr. Run R1's northern outflow exhibits bifurcated lobes due to a reorientation of the outflow during the simulation.
\\
We also compare in Fig. \ref{fig:inandout}b the specific angular momentum $j_{\mathrm{shell}}$ carried by the flow of material, measured as
\begin{equation}
    j_{\mathrm{shell}} = -\frac{R_{\mathrm{shell}}\int_{t} \int_{S} \rho v_{\mathrm{r}}v_{\mathrm{\phi}}dSdt}{\int_{t}\int_{S}\dot{M}dt},
\end{equation}
where $v_{\mathrm{\phi}}$ is the azimuthal velocity. We see that outflows carry 3.85 times more specific angular momentum than the inflows for R1, and 3.41 times for R2. This illustrates the role of magnetic fields in carrying angular momentum outward. However, other transport processes within the disk must be at play to remove the remaining angular momentum, as the outflow does not have sufficient mass to be efficient in this regard.

\subsection{Vertical and radial accretion}
\label{sec:cylmf}

\begin{figure*}[h]
\centering
\includegraphics[scale=.25]{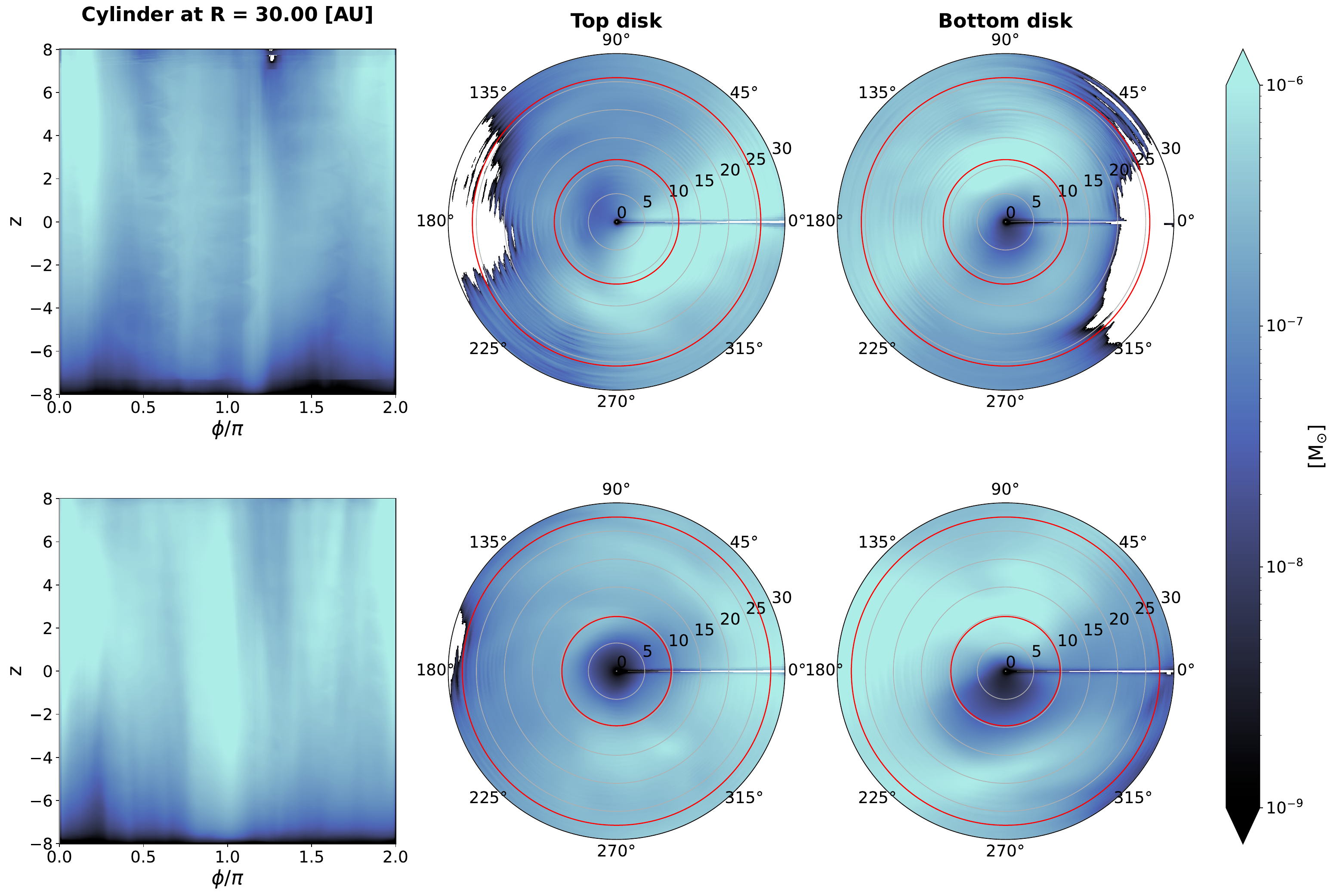} 
\caption{Total inflow of material through a cylinder of radius $R_{\mathrm{shell}}=~30$~AU and height $h=16$~AU throughout the simulation's duration post-sink formation for run R1 ($\approx 58$ kyr, first row) and R2 ($\approx 27$ kyr, second row), computed by integrating Eq. \ref{eq:mdotcyl} (first column) and Eq. \ref{eq:mdotz} (second and third columns) in time. The red circles in the second and third columns correspond to the minimum and maximum semi-major axis of the disk throughout each simulation's duration, and thus represent the locations where material has landed in the disk.}
\label{fig:cylacc}
\end{figure*}

\begin{figure}[h]
\centering
\includegraphics[scale=.3]{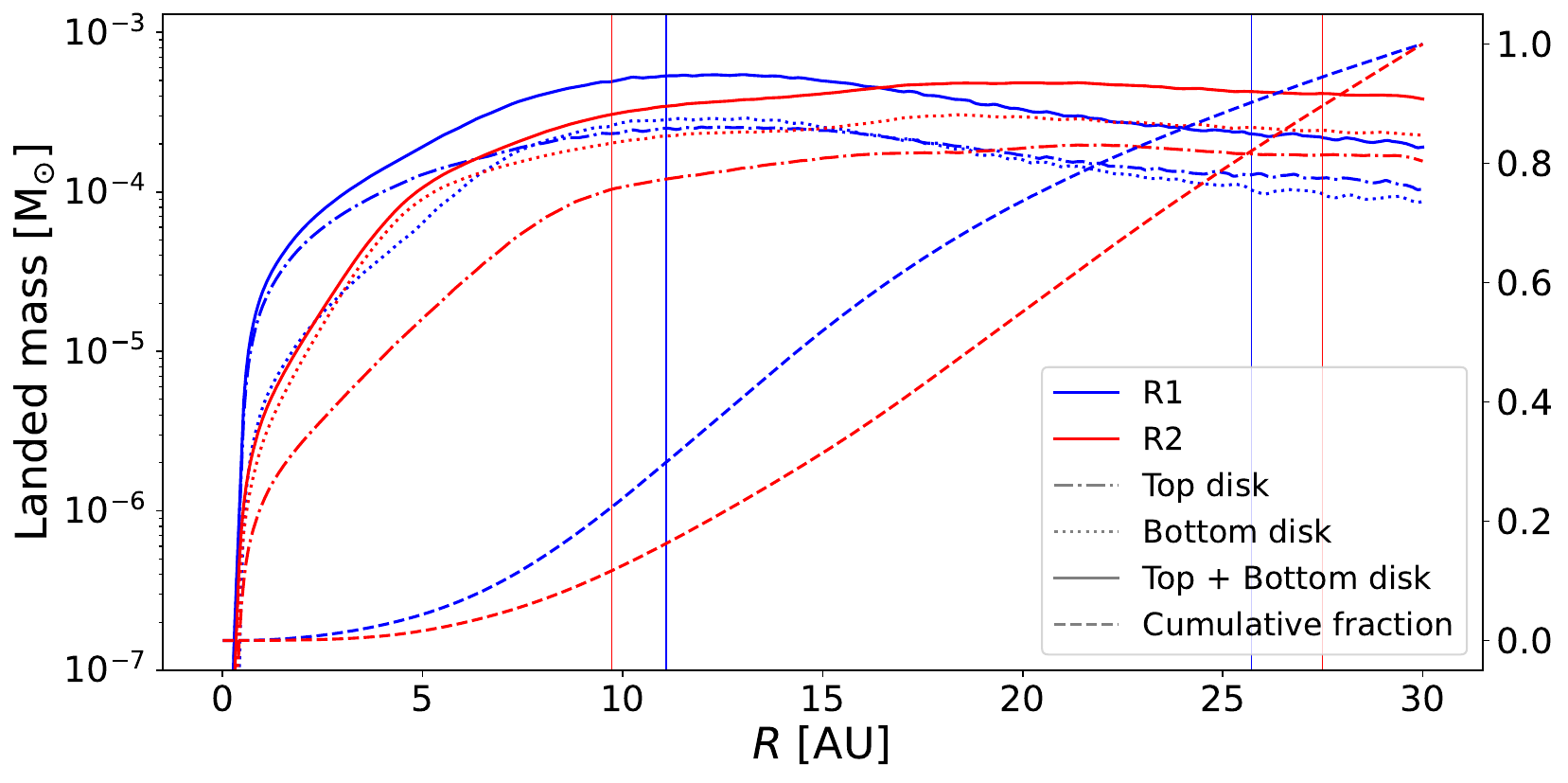} 
\caption{Vertical mass accretion onto the disk for runs R1 (blue) and R2 (red), obtained by integrating Eq. \ref{eq:mdotz} over $\phi$ and time. Dash-dotted and dotted lines show accretion through the top ($z=8$ AU) and bottom ($z=-8$ AU) surfaces, respectively; the solid lines are their sum. Solid vertical lines mark the disk’s minimum and maximum radii during the simulation. The dashed lines represent the cumulative fraction of landed mass (normalized to unity), plotted against the right-hand axis. This figure is complementary to Fig. \ref{fig:cylacc}.}
\label{fig:topandbottom}
\end{figure}

\begin{figure}[h]
\centering
\includegraphics[scale=.42]{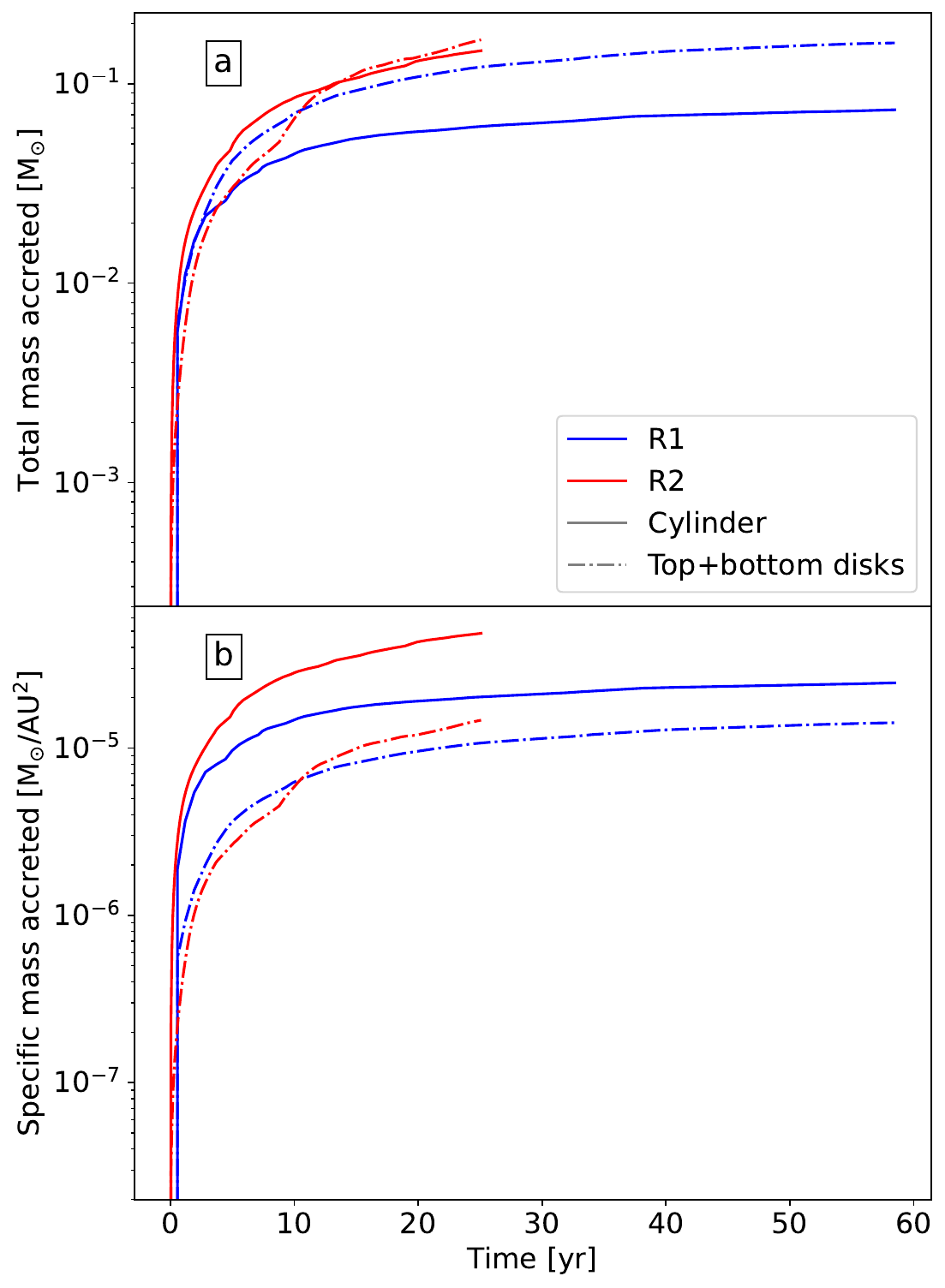} 
\caption{A comparison of the total (panel a) and surface specific (panel b) mass accreted through the cylinder of radius $R_{\mathrm{cyl}}~=~30$~AU (solid lines), and the top and bottom disks of the cylinder at respectively $z=8$ AU and $z=-8$ AU (dash-dotted lines). These are computed by integrating Eqs. \ref{eq:mdotcyl} and \ref{eq:mdotz} in space and time.}
\label{fig:verticalvsradialacc}
\end{figure}

Once material crosses the inner 30 AU radius, we examine how it reaches the disk by computing mass fluxes across a fixed cylindrical surface with $R_{\mathrm{cyl}} = 30$ AU and height $h_{\mathrm{cyl}} = 16$ AU. This allows a comparison of the radial inflow through the cylinder wall and vertical inflow through its top and bottom surfaces at $z = \pm 8$ AU. We emphasize however, that this cylinder does not represent the disk surface, but rather acts as its proxy. The radial mass accretion rate is (locally)
\begin{equation}
\label{eq:mdotcyl}
    \dot{M}_{\mathrm{cyl}}(z, \phi, t) = -\rho v_{\mathrm{R}}R_{\mathrm{cyl}}d\phi dz,
\end{equation}
where $v_{\mathrm{R}}$ is the radial velocity in cylindrical coordinates and $\phi$ the azimuth. The vertical accretion rate through the top and bottom surfaces is (locally)
\begin{equation}
\label{eq:mdotz}
    \dot{M}_{\mathrm{z}}(R, \phi, t) = -\mathrm{sgn}(z)\rho v_{\mathrm{z}}RdR d\phi,
\end{equation}
where $\mathrm{sgn}$ is the sign function, $v_{\mathrm{z}}$ the vertical velocity, and $R=\sqrt{x^2+y^2}$.
\\
Figure~\ref{fig:cylacc} shows the time integrated mass fluxes. Radial accretion is anisotropic in $\phi$ but less so across $z$. Vertical accretion at $z = \pm 8$ AU (second and third columns) reveals negative values in R1 from current sheets that vertically recirculate material\footnote{See Appendix \ref{appendix:EOaccr} for details.}, with mass re-injected elsewhere radially or through the opposite face. Red lines mark the minimum and maximum disk radius over time. Both runs show substantial accretion in the inner disk, with some material landing at radii as small as $\sim 10$ AU.
\\
In Fig.~\ref{fig:topandbottom}, the azimuthally and temporally integrated vertical accretion $\int_{t}\int_{\phi}\dot{M}_{\mathrm{z}}\,dt$ indicates that vertical inflow delivers mass fairly evenly across disk radii\footnote{We note an apparent asymmetry between top and bottom inflow, which is likely circumstantial given the identical initial conditions but differing cloud masses.}. 
\\
Figures~\ref{fig:verticalvsradialacc}a,b compare the total and surface-specific accretion from radial (solid lines) and vertical (dash-dotted lines) directions. In R1, most mass arrives vertically, while in R2, radial and vertical contributions are similar. Nevertheless, radial accretion dominates per unit area over vertical accretion in both runs. R2’s enhanced radial inflow stems from its greater angular momentum budget, which concentrates more material near the midplane (see Fig.~\ref{fig:coldens}c,f).
\\
In contrast with classical 1D hydrodynamic models, where material falls at its centrifugal radius, our simulations reveal that magnetic fields, in concert with turbulence, play an important role by rendering accretion highly anisotropic.

\subsection{Tracer particles}

\begin{figure}[h]
\centering
\includegraphics[scale=.45]{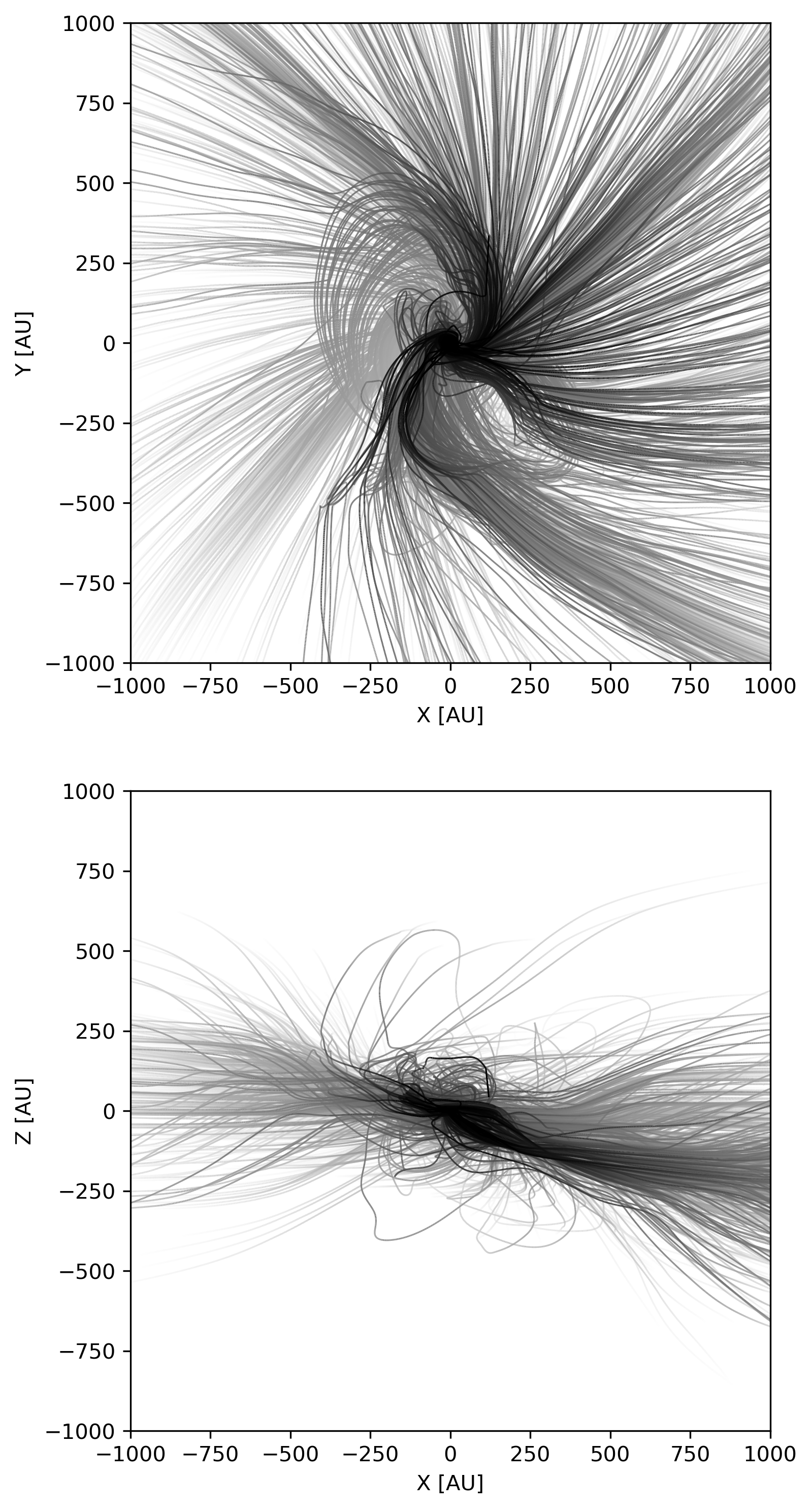} 
\caption{Trajectories of the 5645 tracer particles that settle into the disk during run R2 in the $x$–$y$ (top) and $x$–$z$ (bottom) planes. The color scale reflects the cumulative distance traveled by each particle, with darker shades marking the most recent segments of each path.}
\label{fig:tracertraj}
\end{figure}

\begin{figure}[h]
\centering
\includegraphics[scale=.4]{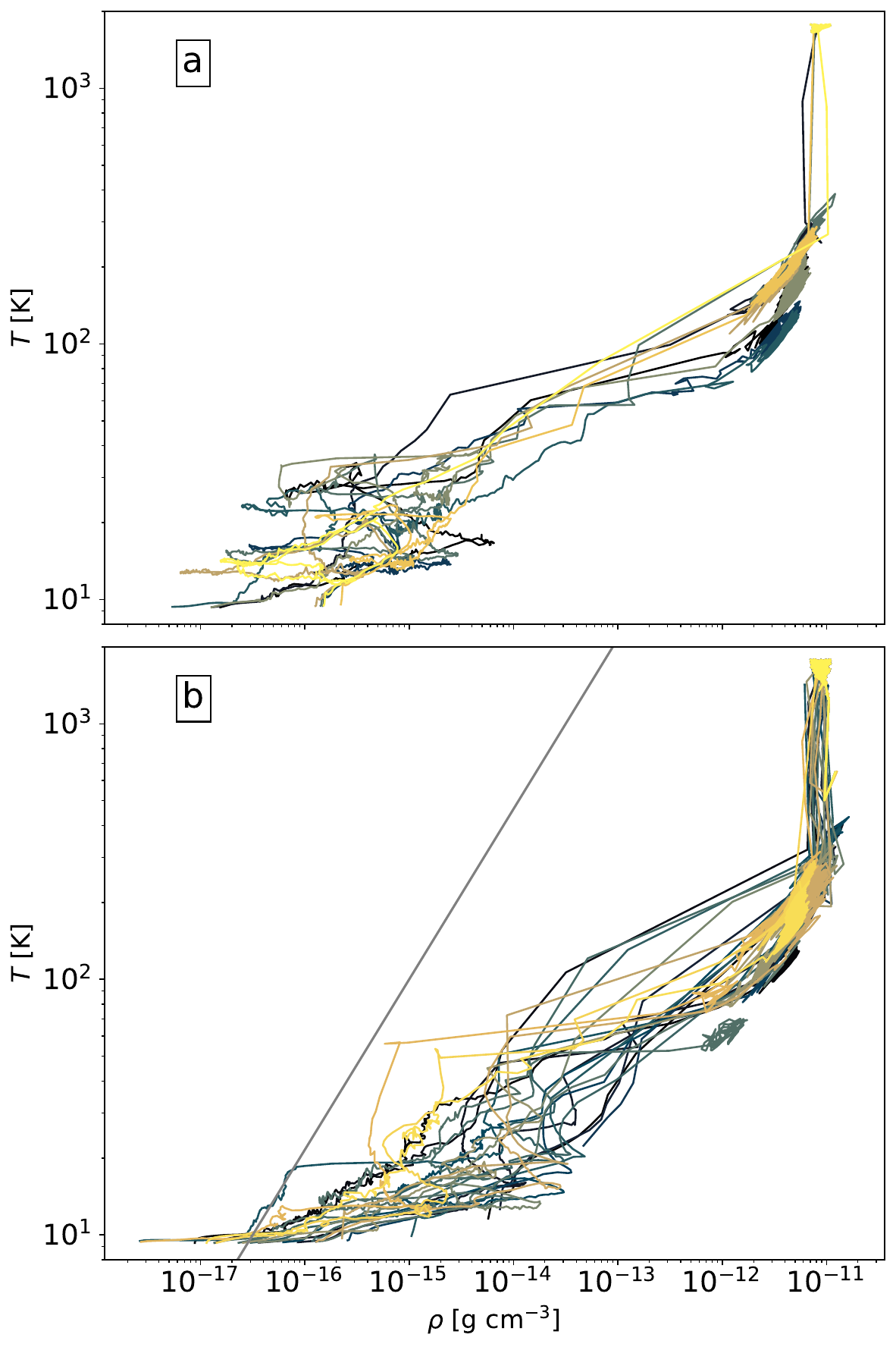} 
\caption{Temperature–density histories of randomly selected tracer particles that settle onto the disk from the polar regions (top, panel a) and from near the equatorial plane (bottom, panel b). Each curve corresponds to an individual particle, represented by a different color. The gray curve in panel (b) represents $T\propto\rho^{5/3 -1}$.}
\label{fig:tracerhist}
\end{figure}

Run R2 employs $10^{4}$ gas tracer particles, of which 5645 are located within the disk or the sink accretion volume by the final simulation snapshot. Figure~\ref{fig:tracertraj} shows the trajectories of these particles in the $x$–$y$ (top) and $z$–$x$ (bottom) planes. These trajectories are consistent with our earlier results, confirming that the majority of the accreted mass arrives onto the disk near the equatorial plane. In the $x$–$y$ view, particles clearly trace the prominent interchange instability, with distinct streamers feeding material into the disk.  These trajectories arise naturally from the particle–mesh method, which drives tracers toward regions of higher density\footnote{This method also drives them toward regions of converging flows.}. The $z$–$x$ view highlights the highly irregular and chaotic nature of vertical accretion, with a wide range of particle trajectories in this direction. Unfortunately, once the particles have settled in the disk, our snapshot frequency does not permit us to follow their evolution within it. In addition, most of these particles are ultimately accreted by the sink particle early in the disk's history, as they migrate inward during phases of inward-driven transport (see Sec.~\ref{sec:masstransport}) and can no longer escape its immediate vicinity.
\\
Having established the spatial trajectories of the tracers, we now turn to their thermodynamic histories during infall. Figure~\ref{fig:tracerhist} shows the temperature–density histories of randomly selected tracer particles that settle onto the disk either from the polar regions or from near the equatorial plane. Particles accreted through the equatorial plane (Fig. \ref{fig:tracerhist}b) exhibit a more classical adiabatic heating (gray curve) trajectory prior to disk settling, with a sharp density increase once they settle onto the disk. By contrast, particles arriving from the polar regions (Fig. \ref{fig:tracerhist}b) display more complex evolutionary paths, undergoing repeated cooling and heating events prior to disk settling as they traverse different regions of the flow, being influenced by both the magnetic field and the outflow.

\section{Disk kinematics}

The anisotropic nature of accretion imprints itself on the disk’s velocity field. Rather than settling into purely circular Keplerian orbits, the localized streams of infalling gas produce eccentric motions \citep{commercon_2024}, and as we will show below, drive mass transport in the interior. In this section, we explore these kinematic features and how they shape the disk’s evolution.

\subsection{Eccentric motions within disk}
\label{sec:e}   

\begin{figure}[h]
\centering
\includegraphics[scale=.3]{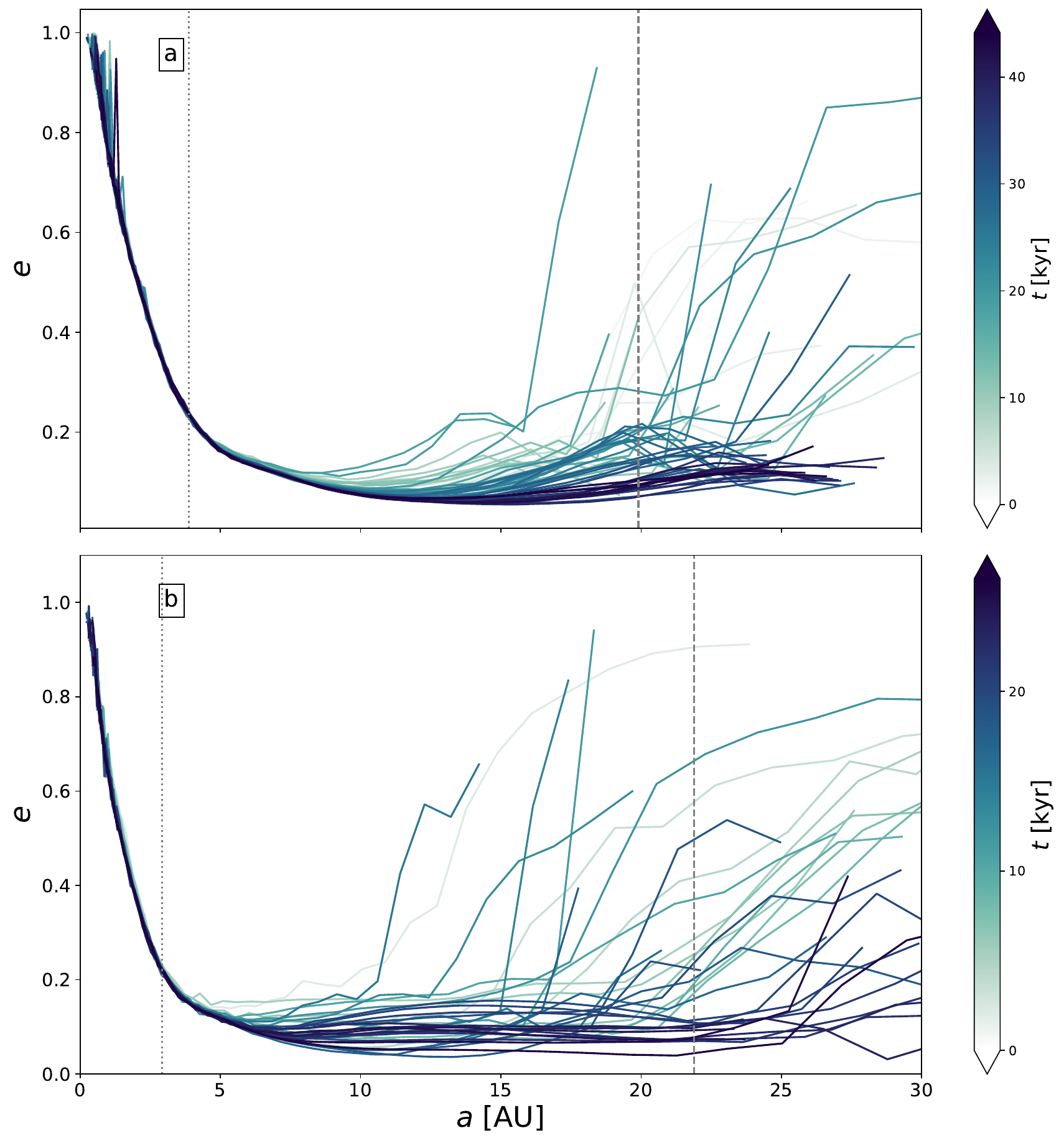} 
\caption{Distribution of eccentricity, computed from the disk kinematics (Eqs. \ref{eq:e}, \ref{eq:a}), for run R1 (a) and run R2 (b). Each curve represents a different time, with $t=0$ marking the epoch of sink formation. The dashed line indicates the largest radius enclosing 90\% of the disk mass over the entire simulation. The sink accretion radius is indicated by the dotted line.}
\label{fig:kinE}
\end{figure}

\begin{figure}[h]
\centering
\includegraphics[scale=.5]{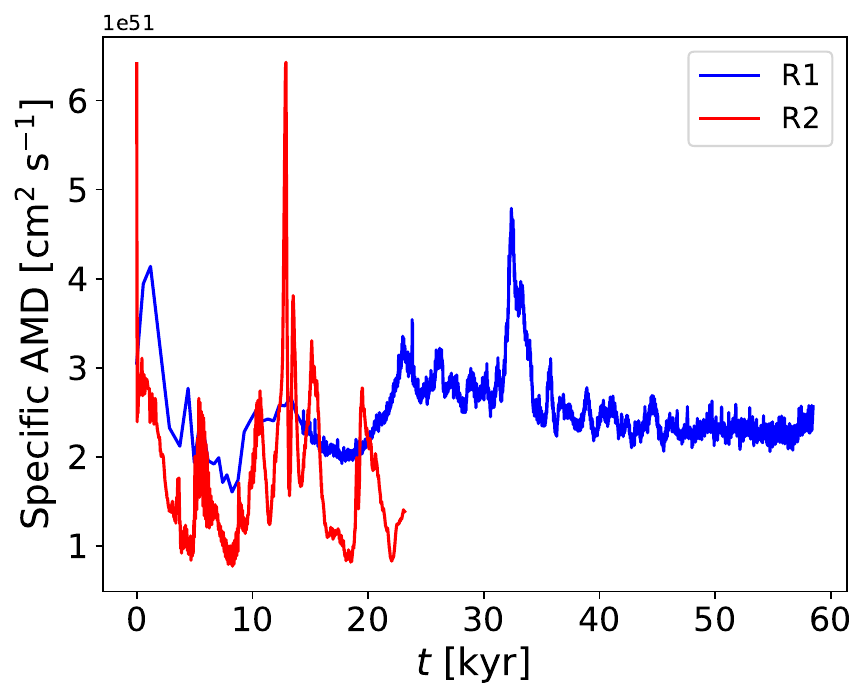} 
\caption{Specific angular momentum deficit within the disk as a function of time since sink formation for runs R1 (blue) and R2 (red).}
\label{fig:AMD}
\end{figure}

As noted earlier, the disks in our simulations exhibit ellipsoidal morphologies, with apparent eccentricities often exceeding $\sim 0.1$ (Fig.~\ref{fig:globalprops}c). We quantify eccentric motions via the fluid kinematics following \cite{commercon_2024}:
\begin{equation}
    \label{eq:e}
    \vec{e} = \frac{\vec{v}\times\vec{j}}{GM_{*}} - \frac{\vec{r}}{r},
\end{equation}
where $\vec{e}$ is the eccentricity vector and $r=\sqrt{x^{2}+y^{2}+z^{2}}$. $M_{*}$ is the sink (i.e., protostellar) mass. The semi-major axis of each orbit is then obtained as
\begin{equation}
    \label{eq:a}
    a = \left ( \frac{2}{r}-\frac{v^{2}}{GM_{*}}\right )^{-1}.
\end{equation}
To account for pressure contributions on the orbital velocity, we measure the average eccentricity within an orbital bin. The resulting values are presented in Fig. \ref{fig:kinE}. We find that disk eccentricity is progressively damped over time. In run R2 (Fig.~\ref{fig:kinE}b), episodes of enhanced accretion (Fig.~\ref{fig:globalprops}a) coincide with temporary eccentricity growth, followed by rapid damping. Elevated eccentricities persist in the inner disk, probably from the influence of the sink particle. To assess this, we compute the angular momentum deficit (AMD) within the disk, which represents  the difference between the circular orbit angular momentum and the actual orbital angular momentum:
\begin{equation}
    \mathrm{AMD} = \int_{a}\int_{0}^{2\pi}\Sigma(a)\sqrt{a^{3}GM_{*}}\left ( 1 - \sqrt{1-e^{2}} \right ) dad\phi,
\end{equation}
where $\Sigma(a)$ is the mean column density within an orbital bin. In the limit of an ideal, inviscid disk, AMD is a conserved quantity. The resulting specific AMD (Fig. \ref{fig:AMD}) shows a general decrease over time for R1, whereas the trend is not as clear for R2. It also shows increases during strong accretion episodes. We attribute the initial ($t<10$~kyr) declining trend to grid dissipation, as it is stronger when the disk is small and the orbital resolution is inadequate. The declining trend weakens for R1 when the disk grows in radius and numerical resolution becomes sufficient. 
In the quiescent regime of run R1 ($t>35$~kyr), we can estimate a characteristic damping timescale under the assumption that the disk turbulence acts as an effective bulk viscosity \citep{goodchild_2006}. This timescale is comparable to the viscous timescale, approximated as $(\alpha\Omega)^{-1}(R/H)^{2}$. Based on the disk properties inferred from Fig. \ref{fig:globalprops}b ($M_{*}\sim 0.3~\mathrm{M_{\odot}}$, $R_{\mathrm{d}}\sim 15$~AU, $H/R \sim 0.1$) and the effective viscosity $\alpha \sim 0.1$,\footnote{See Sec. \ref{sec:masstransport}.} we estimate this timescale to be $\sim 17$~kyr. While the exact dissipative properties of the turbulence remain complex \citep{ogilvie_2001}, this estimate highlights that physical damping timescales are short enough to be relevant within the remaining $25$~kyr simulation time. Thus, the damping we observe may not be purely numerical. Consequently, the sustained eccentricity implies that anisotropic accretion actively replenishes AMD against these dissipative processes.
\\
While we observe significant eccentricity in the Class 0 phase driven by anisotropic infall, this state is likely transient. The existence of the Cold Classical population in the Kuiper Belt, which formed during the later Class II phase and exhibits quasi-circular orbits, suggests that the disk eventually circularizes. This provides further evidence that AMD is not conserved; once accretion subsides, damping reduces the eccentricity, leading to the dynamically cold disks in which planetary bodies eventually form.

\subsection{Orbital mass transport}
\label{sec:masstransport}
\begin{figure*}[h]
\centering
\includegraphics[scale=.26]{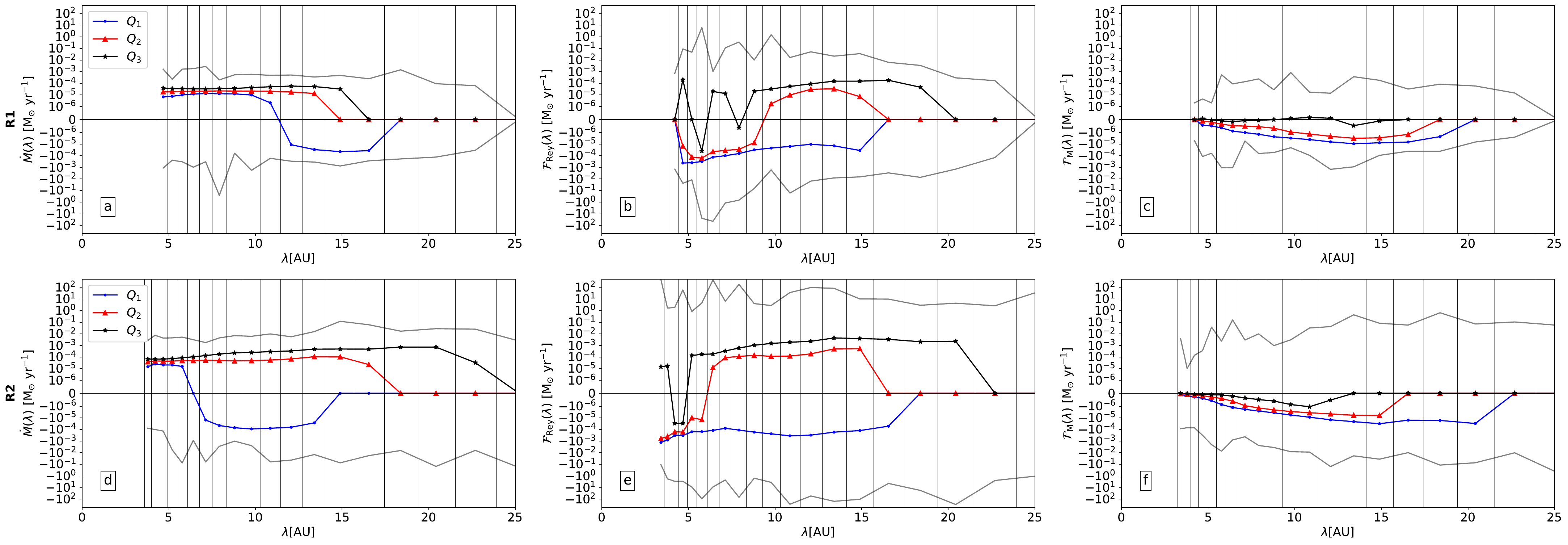} 
\caption{The internal kinematics of the disk. Shown here are orbital mass accretion rate (a, d), turbulent stress tensor induced accretion rate (b, e), and Maxwell stress tensor induced accretion rate (c, f) for runs R1 (first row) and R2 (second row), computed as described in Sec. \ref{sec:masstransport}. These are computed in fixed logarithmic bins in $\lambda$ (vertical lines) throughout all simulation snapshots using only cells belonging to the disk, and we display the resulting first (blue), second (red), and third (black) quantiles, which respectively represent the 25th, 50th, and 75th percentiles. The gray curves in each plot represents minimal and maximal values.}
\label{fig:internaltransport}
\end{figure*}

\begin{figure}[h]
\centering
\includegraphics[scale=.26]{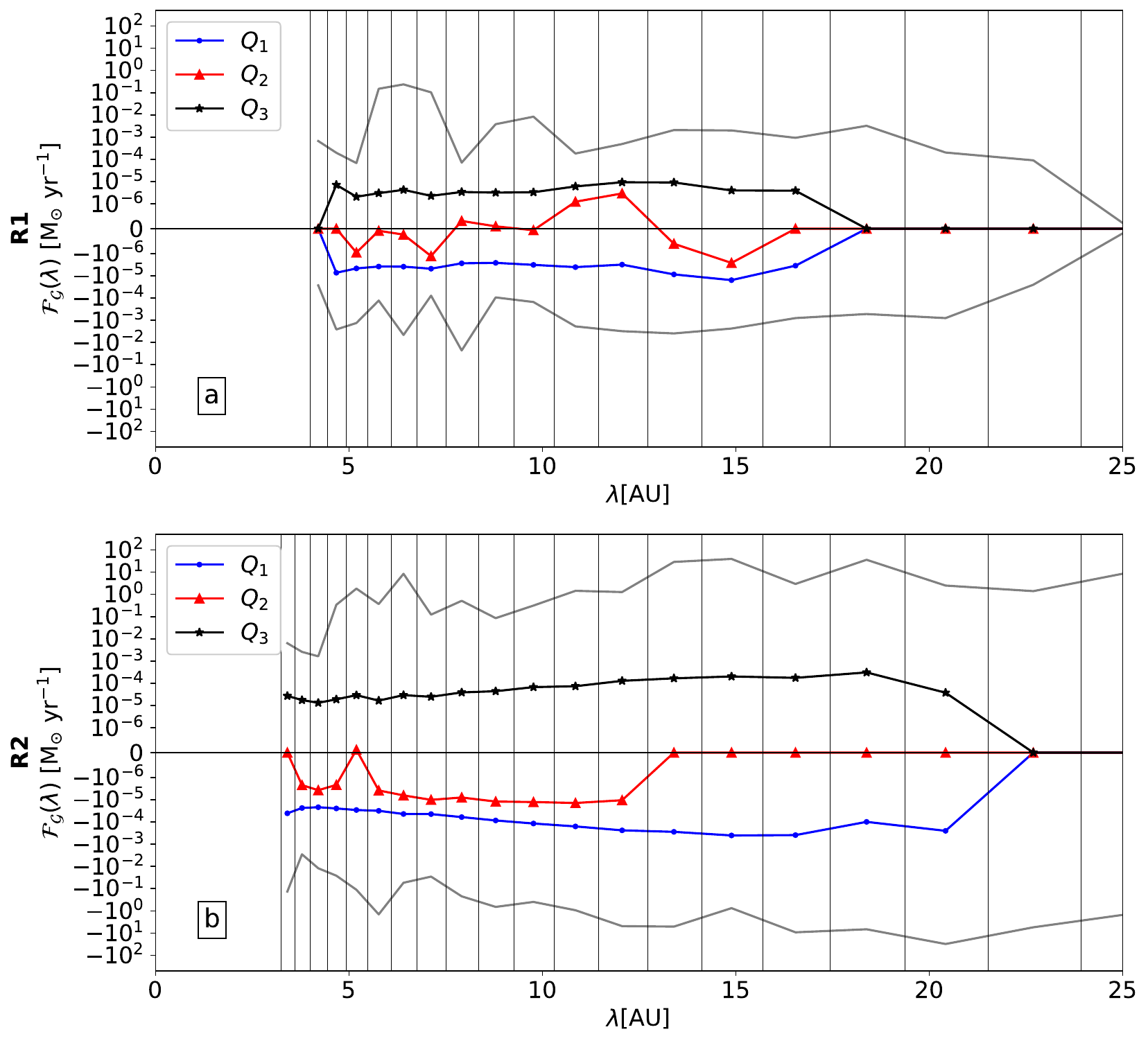} 
\caption{Same as Fig. \ref{fig:internaltransport}, but this time showing the gravitational stress tensor induced accretion rate $\mathcal{F}_{\mathrm{\mathcal{G}}}$.}
\label{fig:gravtransport}
\end{figure}

\begin{figure}[h]
\centering
\includegraphics[scale=.26]{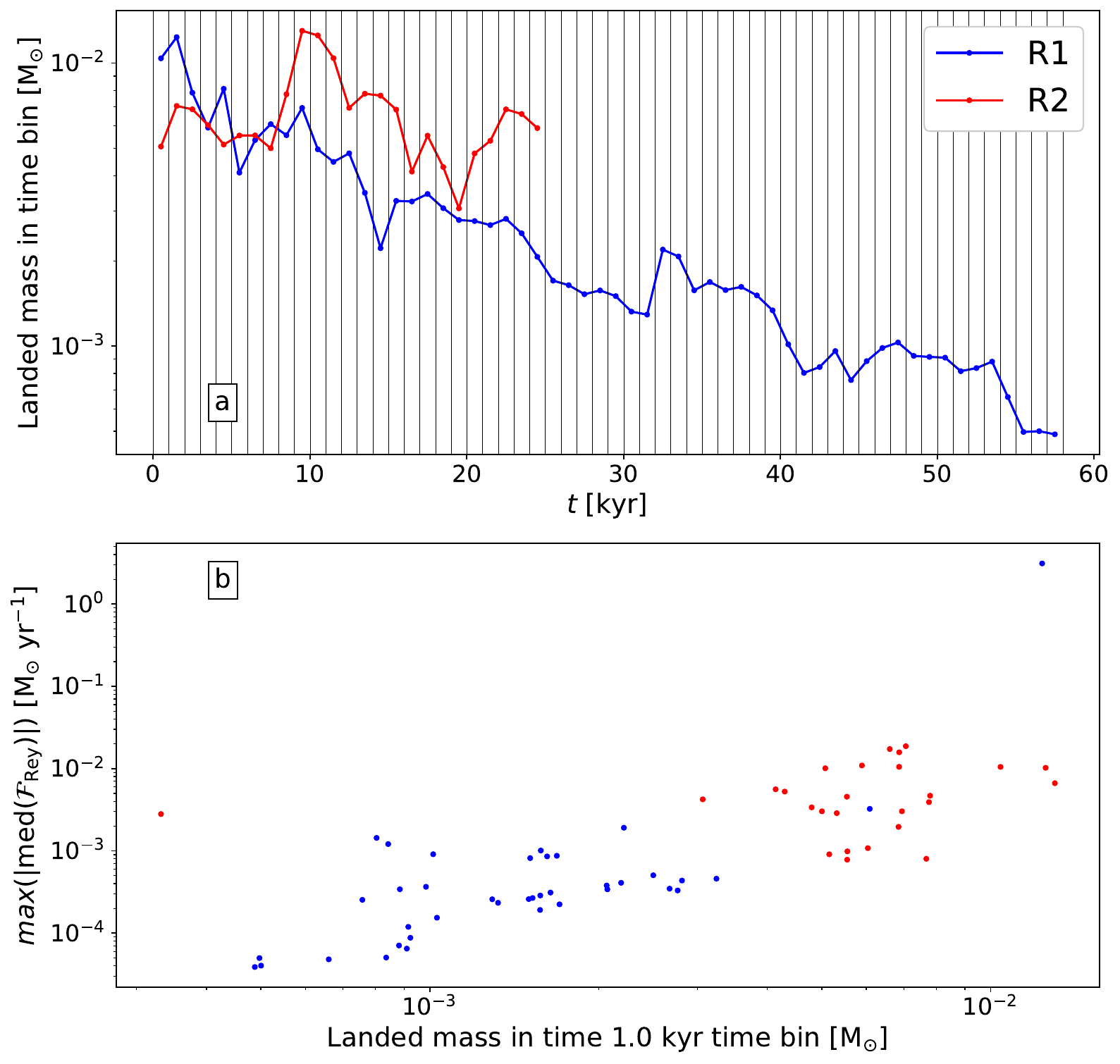} 
\caption{Correlation between vertical accretion and turbulent stress in runs R1 (blue) and R2 (red). Panel (a) displays the mass accreted vertically onto the disk in 1 kyr bins (vertical lines). Panel (b) shows the maximum median accretion rate within each bin, plotted against the corresponding accreted mass.}
\label{fig:correlation}
\end{figure}

Eccentricity is only one aspect of the disks’ dynamics, as the inflow patterns seen in Fig.~\ref{fig:topandbottom} also suggest vigorous vertical mass accretion across all radii. Such accretion can drive turbulence \citep{klessen_hennebelle}, which in turn causes angular momentum transport through the turbulent stress tensor \citep{balbus_1999}.
\\
Assessing mass transport in cylindrical or spherical coordinates is challenging, as eccentric fluid motions bias the measured fluxes. To account for the disk’s nested eccentric orbits, we adopt the orbital formalism of \cite{ogilvie_2014}, which uses $(\lambda, \phi, z)$ coordinates, where the semi-latus rectum
\begin{equation}
    \lambda = a(1-e^2) = \frac{j^2}{GM_{*}}
\end{equation}
is conserved within an orbit if angular momentum is conserved, making it a convenient quasi-radial coordinate. We therefore measure mass transport in $\lambda$ bins via
\begin{equation}
    \label{eq:orbitalmacc}
    \dot{M}(\lambda) = \int\int J\rho \left ( \frac{v_{\mathrm{r}}}{R_{\lambda}} - \frac{R_{\phi}}{R_{\lambda}}\Omega \right ) d\phi dz,
\end{equation}
where $J=\partial (x,y,z)/\partial (\lambda,\phi,z)$ is the Jacobian of the coordinate system (see \citealp{ogilvie_2014})\footnote{See Appendix B of \citealp{ogilvie_2014} for the mathematical expressions of $R_{\lambda}$ and $R_{\phi}$.}.
\\
Turbulent stresses will induce transport
\begin{equation}
\label{eq:reynoldsmacc}
    \mathcal{F}_{\mathrm{Rey}}(\lambda)=-\frac{\partial \mathcal{G}/\partial \lambda }{dj/d\lambda },
\end{equation}
where the subscript stands for 'Reynolds' and $\mathcal{G}$ is the internal torque computed using
\begin{equation}
    \mathcal{G} = -\int \int JR^{2}T^{\lambda\phi}d\phi dz,
\end{equation}
and $T^{\lambda \phi}$ is a component of the turbulent stress tensor
\begin{equation}
    T^{\lambda\phi} = \frac{R}{J} T^{R\phi}-\frac{RR_{\phi}}{J}T^{\phi\phi},
\end{equation}
where
\begin{equation}
    T^{R\phi} = \delta v_{\mathrm{r}}\delta v_{\phi},
\end{equation}
\begin{equation}
    T^{\phi\phi} = \delta v_{\phi}^{2}.
\end{equation}
Here, $\delta v_{\mathrm{r}}$ and $\delta v_{\phi}$ are deviations from Keplerian orbital velocity in the radial and azimuthal components, respectively\footnote{They are contra-variant components.}. 
\\
Magnetic fields also induce transport through the Maxwell stress tensor:
\begin{equation}
    T^{R\phi}_{\mathrm{M}} = v_{\mathrm{A,R}}v_{\mathrm{A,\phi}},
\end{equation}
\begin{equation}
    T^{\phi\phi}_{\mathrm{M}} = v_{\mathrm{A,\phi}}^{2},
\end{equation}
where $\vec{v}_{\mathrm{A}} = \vec{B}/\sqrt{4\pi\rho}$ is the Alfvén speed. These components yield a Maxwell torque and corresponding internal mass flux $\mathcal{F}_{\mathrm{M}}(\lambda)$. Finally, we measure a gravitational stress tensor induced mass flux $\mathcal{F}_{\mathrm{\mathcal{G}}}$ using
\begin{equation}
    T^{R\phi}_{\mathrm{\mathcal{G}}} = v_{\mathrm{g,R}}v_{\mathrm{g,\phi}},,
\end{equation}
\begin{equation}
    T^{\phi\phi}_{\mathrm{\mathcal{G}}} = v_{\mathrm{g,\phi}}^{2},
\end{equation}
where $\vec{v_{g}}=-\vec{g}/\sqrt{4\pi G\rho}$ and $\vec{g}$ is the gravitational acceleration.
\\
Note here that positive values of $\dot{M}$, $\mathcal{F}_{\mathrm{Rey}}$, $\mathcal{F}_{\mathrm{M}}$ and $\mathcal{F}_{\mathrm{\mathcal{G}}}$ correspond to decretion,  i.e., outward mass transport in the $\lambda$ direction.
\\
As accretion is highly time-dependent and non-monotonic, measuring mass transport rates from individual snapshots provides limited insight into the cumulative contributions of different terms over the simulation. We therefore perform a statistical analysis of the orbital mass transport. Fixed logarithmic bins in $\lambda$ are defined, within which we compute $\dot{M}(\lambda)$, $\mathcal{F}_{\mathrm{Rey}}$, $\mathcal{F}_{\mathrm{M}}$ and $\mathcal{F}_{\mathrm{\mathcal{G}}}$ for each snapshot using only cells belonging to the disk. All quantities are vertically averaged with density weighting, thus biasing the results with the dense midplane. By this stage of the simulation, the maximum refinement level is reached and time between snapshots becomes roughly constant. The statistical distributions are shown in Fig.~\ref{fig:internaltransport} (for $\dot{M}(\lambda)$, $\mathcal{F}_{\mathrm{Rey}}$, and $\mathcal{F}_{\mathrm{M}}$) and Fig.~\ref{fig:gravtransport} (for $\mathcal{F}_{\mathrm{\mathcal{G}}}$), where we plot the zeroth (Q0, minimum, bottom gray curves), first (Q1, 25th percentile, blue curves), second (Q2, median, red curves), third (Q3, 75th percentile, black curves), and fourth (Q4, maximum, top gray curves) quantiles for each $\lambda$ bin, for run R1 (top rows) and R2 (bottom rows). The innermost regions strongly influenced by the sink particle are excluded from the plot.
\\
The orbital mass transport profiles in Fig.~\ref{fig:internaltransport}a,d show median $\dot{M}$ values of $\sim 10^{-5}~\mathrm{M_{\odot}\ yr^{-1}}$ for both runs, indicating that material mainly spreads outward in the midplane. Beyond $\lambda \approx 14$ AU for R1 and $\lambda \approx 18$ AU for R2, radial motions are negligible, as the magnetic field truncates the disk \citep{lee_2021}. However, Q1 curves (blue) reveal intermittent inward transport, with a larger spread in R2 due to its higher disk mass. The turbulent stress tensor (Fig.~\ref{fig:internaltransport}b,e) drives outward transport in the outer disk ($\lambda \gtrsim 8~\mathrm{AU}$ for R1; $\lambda \gtrsim 6~\mathrm{AU}$ for R2) and inward transport in the inner regions, with median rates of $\sim 10^{-5}~\mathrm{M_{\odot}\ yr^{-1}}$ for R1 and $\sim 10^{-4}~\mathrm{M_{\odot}\ yr^{-1}}$ for R2. By contrast, the Maxwell stress (Fig.~\ref{fig:internaltransport}c,f) consistently drives inward transport, but at much lower amplitudes (median rates of $\sim 10^{-6}~\mathrm{M_{\odot}\ yr^{-1}}$ for R1, $\sim 10^{-5}~\mathrm{M_{\odot}\ yr^{-1}}$ for R2), rendering it largely negligible compared to turbulent stresses. $\mathcal{F}_{\mathrm{\mathcal{G}}}$ displays more complex behavior in run R1 (Fig. \ref{fig:gravtransport}a), with median accretion rates oscillating between positive and negative values, albeit at negligible amplitudes when compared to turbulent and Maxwell stress. The more dynamical run R2 has $\mathcal{F}_{\mathrm{\mathcal{G}}}$ display negative median rates (Fig. \ref{fig:gravtransport}b), indicating inward transport, albeit once again at negligible rates ($\sim 10^{-5}~\mathrm{M_{\odot}~yr^{-1}}$) when compared to turbulent stresses. The disks are gravitationally stable (Appendix \ref{appendix:toomre}) as they are hot, which explains the negligible gravitational stress.
\\
Notably, in the inner disk, $\dot{M}(\lambda)$ remains positive even when $\mathcal{F}_{\mathrm{Rey}}$ is negative, likely due to sink particle effects and limited numerical resolution.
\\
Overall, Fig.~\ref{fig:internaltransport} indicates that midplane transport is dominated by turbulent stress, leading predominantly to outward spreading. For a typical disk with scale height $\approx 2.5~\mathrm{AU}$, sound speed $\approx 10^{5}~\mathrm{cm\ s^{-1}}$, and column density $\approx 10^{2}~\mathrm{g\ cm^{-2}}$, the measured median outward mass accretion rate of $\sim 10^{-5}~\mathrm{M_{\odot}\ yr^{-1}}$ corresponds to an effective \cite{shakura_1973} viscosity of $\alpha_\mathrm{sh} \approx 0.09$ (with $\alpha_\mathrm{sh}=\dot{M}/3\pi c_\mathrm{s} h \Sigma$). We stress, however, that an $\alpha$-disk prescription is not applicable in this context, as the turbulent transport exhibits strong temporal and spatial variability.
\\
A key question remaining is the origin of this turbulent stress. To test whether accretion drives turbulence, we measure the amount of vertically landed mass in 1 kyr bins, which is long enough for several orbits in the outer disk and dozens in the inner disk. In Fig. \ref{fig:correlation}a, the vertically accreted mass declines from $10^{-2}$ to $10^{-3}\ \mathrm{M_{\odot}}$ in R1, while R2 shows a more variable but nevertheless decreasing trend. Fig. \ref{fig:correlation}b applies the same statistical method as Fig. \ref{fig:internaltransport}, but now takes the maximum absolute median accretion rate across orbits in each temporal bin and plots it against the corresponding landed mass. The strong correlation confirms a link between vertical accretion and internal transport through turbulent stress.

\section{Discussions} \label{section:discussions}

\subsection{Asymmetric feeding of the disk}

As in \cite{commercon_2024}, we have shown that the interplay between magnetic fields and turbulence produces anisotropic inflows in the form of streamers onto the disk. These streamers excite and sustain eccentric motions within the disk by continuously supplying material with an angular-momentum deficit. This provides a natural explanation for eccentric protoplanetary disks without requiring extended spiral waves or the presence of planets or stellar companions. Robust models like those presented here may be used to account for the complex kinematic features observed in disks (e.g., \citealp{zamponi_2021, thieme_2023, vanthoff_2023, winter_2025}).
\\
Perhaps the most novel result of this study is that asymmetric accretion also enables material to land vertically onto the disk, thereby driving turbulence within it. This turbulence leads to efficient angular-momentum transport inside the disk, which dominates over contributions from the Maxwell and gravitational stress. 
Thus, while magnetic fields govern angular-momentum transport outside the disk, the asymmetric feeding they facilitate becomes its primary driver within it. Most notably, this process yields highly efficient transport, equivalent to an effective viscosity parameter of $\alpha_\mathrm{sh}=\dot{M}/3\pi c_\mathrm{s} h \Sigma \approx 0.1$, without the need to invoke magnetic instabilities such as the magneto-rotational instability (MRI; \citealp{balbus_1991}).

\subsection{A link with cosmochemistry}

\begin{figure}[h]
\centering
\includegraphics[scale=.25]{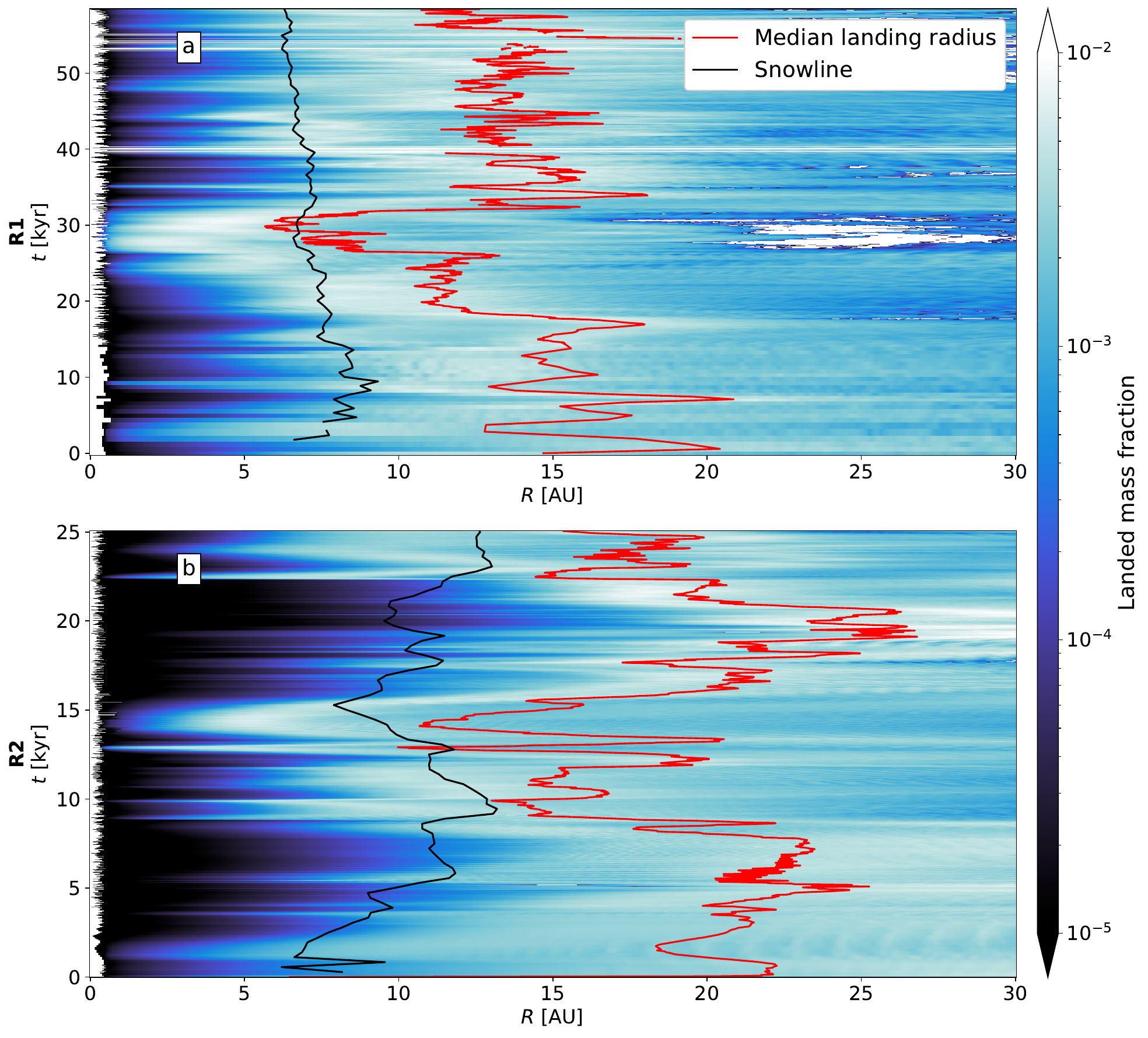} 
\caption{Temporal evolution of the fraction of mass landing at each radius, obtained by normalizing the incoming mass flux from Eq. \ref{eq:mdotz} for each snapshot, for run R1 (a) and run R2 (b). The red curve indicates the median landing radius, where half of the mass lands at larger radii and half at smaller radii. The black curve indicates the location of the snowline. An animated movie, combining panel (b) with a 3D volume render of the disk of run R2 is available.}
\label{fig:landedmassfrac}
\end{figure}

The efficient angular momentum transport driven by turbulence indicates that disk material predominantly spreads outward, with only occasional episodes of inward motion (Fig.~\ref{fig:internaltransport}).
\\
These results are relevant to cosmochemistry, where an isotopic dichotomy is observed in Solar System meteorites \citep{nanne_2019, morbidelli_2024}. Refractory inclusions (CAIs, AOAs), formed at high temperatures near the Sun, are found in the colder outer disk, a distribution generally interpreted as evidence for outward transport. In addition, carbonaceous chondrites (CCs) and non-carbonaceous chondrites (NCs) define two distinct isotopic reservoirs: NCs, depleted in supernova-derived nuclides, dominate the inner Solar System, whereas CCs are linked to the outer regions. Explaining this dichotomy requires models that couple infall history with disk dynamics.
\\
One-dimensional viscous models (e.g., \citealp{pignatale_2018, morbidelli_2022, marschall_2023}) can reproduce outward transport of high-temperature condensates if the centrifugal radius remains small ($< 1$ AU) and if the effective viscosity parameter $\alpha_\mathrm{sh}$ is $\sim 0.1$ throughout disk formation. However, such models process all material through high temperatures, which is inconsistent with the presence of volatiles (e.g., hydrogen bearing species) showing non-solar isotopic compositions in the outer disk.
\\
Our simulations show that outward transport of solids can occur through disk spreading, reaching median decretion rates of $\sim [10^{-5}-10^{-4}]~\mathrm{M_{\odot}\ yr^{-1}}$ thanks to Reynold stresses (with an equivalent $\alpha_\mathrm{sh}\approx 0.1$). Because the sub-AU dynamics are unresolved, the spreading material does not attain CAI-forming temperatures. Redistribution by outflows is also possible \citep{bhandare_2024}, though not modeled here as we do not account for jets. Studies resolving the collapse of the first Larson core find a rapidly expanding disk around the protostar \citep{vaytet_2018, machida_2019, ahmad_2024, ahmad_2025, bhandare_2024}, whose radial spread may be sustained by the vertical accretion-driven turbulence reported here. Said vertical accretion initially delivers material across all radii during the first $\sim$60 kyr (Fig.~\ref{fig:topandbottom}), with a time-variable median landing radius (Fig.~\ref{fig:landedmassfrac}). Although transient surges in accretion toward the inner disk occur (Fig. \ref{fig:landedmassfrac}), our models do not suggest a sustained inner-disk bias during the Class 0 phase. The vertical accretion patterns reported here may explain the presence of non-solar isotopic material in CC regions but not the predominantly NC composition of the inner disk. The isotopic dichotomy may therefore reflect (a) a time-dependent change in the isotopic composition of infalling material \citep{morbidelli_2024}, and (b) a different accretion mechanism with a sustained inner-disk bias in vertical accretion at later times\footnote{The ambient magnetic field outside the disk will weaken as the envelope dissipates, reducing the strength of magnetic streams feeding the disk and potentially favoring the inner-most regions of the disk \citep{lee_2021}.}. Pressure maxima or proto-planets then preserve this distinction by stopping radial mixing.
\\
In summary, our results support a hybrid scenario in which the isotopic dichotomy of Solar System meteorites originates from time-dependent infall composition combined with the later emergence of an inner-disk bias in vertical accretion. This dichotomy is subsequently locked in place by disk structures such as pressure maxima or the birth of a proto-Jupiter.

\subsection{Snowline}

The location of the water snowline in disks is of interest for planetesimal formation (\citealp{stevenson_1988, armitage_2016, schoonenberg_2017, morbidelli_2022}). In 1D models, disks are typically assumed to undergo a gradual buildup over several $10^{5}$ yr, with the water snowline drifting outward as the disk heats, followed by an inward contraction as accretion subsides (e.g., \citealp{pignatale_2018, drazkowska_2018, lichtenberg_2021}). However, the chaotic accretion seen in our models suggests that the snowline likely evolves in a far less smooth and predictable manner.
\\
We locate the water snowline by calculating the water saturation pressure along the disk midplane using the Clausius–Clapeyron relation
\begin{equation}
    P_{\mathrm{sat}}(T) = P_{\mathrm{sat},0}\ e^{-T_{\mathrm{a}}/T},
\end{equation}
where $P_{\mathrm{sat},0} = 1.14\times 10^{13}\ \mathrm{g\ cm^{-1}\ s^{-2}}$ and $T_{\mathrm{a}} = 6062$ K \citep{lichtenegger_1991}. The snowline is defined where the water partial pressure, $X_{\mathrm{H_{2}O}} P$, equals the saturation pressure. Assuming an $\mathrm{H_{2}O}$ abundance of $X_{\mathrm{H_{2}O}} = 5\times10^{-4}$ \citep{harwit_1998}, we identify this location in the disk midplane and plot its evolution over time in Fig. \ref{fig:landedmassfrac} for both runs (black lines). The average midplane density and temperature are shown in Appendix \ref{appendix:avprofilesrhoT}.
\\
The two runs show distinct snowline behaviors. In run R1, the gradual decrease in disk mass and slow outward spreading of its radius (Fig. \ref{fig:globalprops}a,b) cause the snowline to drift inward from $\approx 9$ to $\approx 6$ AU as the disk cools. In contrast, run R2 exhibits rapid variations in disk mass and radius (Fig. \ref{fig:globalprops}a,b), leading to a highly dynamic snowline that follows the disk's mass and radius evolution. Accretion-driven increases in disk density and temperature push the snowline outward, reaching $\approx 13$ AU, which reflects the hotter disk conditions in R2.

\subsection{A dust-enriched disk}

\begin{figure}[h]
\centering
\includegraphics[scale=.3]{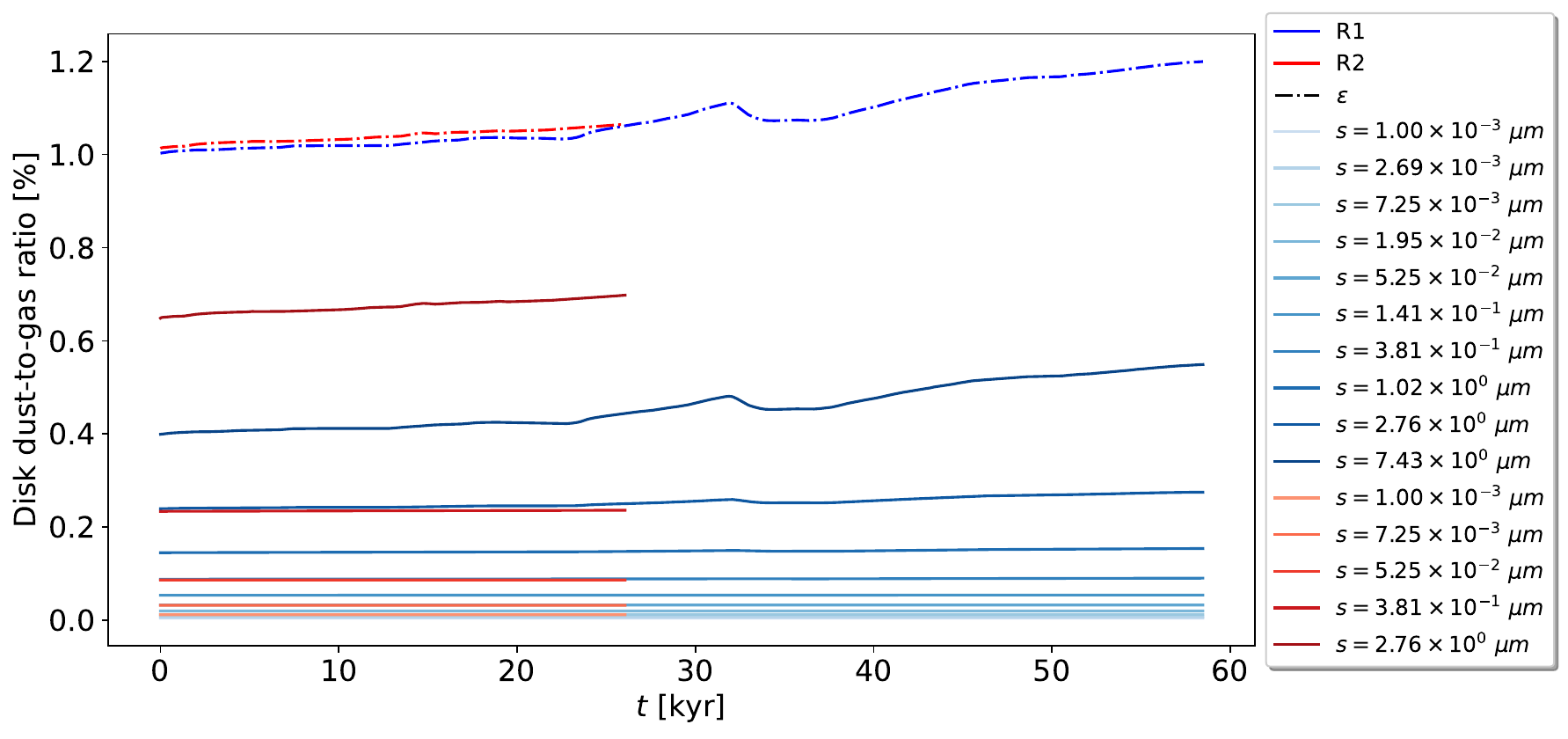} 
\caption{Disk metallicity of runs R1 (blue) and R2 (red) as a function of time, with $t=0$ marking the epoch of sink formation. The dash–dotted line indicates the total dust–to–gas ratio of the disk ($\epsilon$), while the colored solid lines show the ratios of individual dust species, with darker shades corresponding to larger grain sizes.}
\label{fig:eps}
\end{figure}

Having accounted for dust dynamics during the simulation, we may measure the metallicity of the disk and its temporal evolution. We display in Fig. \ref{fig:eps} the dust-to-gas ratio of individual dust species within the disk as a function of time. The total dust-to-gas ratio $\epsilon$ (dash-dotted line) begins at 1\% in both runs, and shows a similar increase over time, with $\epsilon$ reaching $\approx 1.2\%$ by the final snapshot of run R1. However, despite the enrichment, the dust size is too small for appreciable grain growth \citep{krapp_2019, lovascio_2022}.
\\
Once in the disk, all dust species, by virtue of their small size and the high gas density (and hence small Stokes numbers), are tightly coupled to the gas and perfectly follow its kinematics. A more detailed analysis of dust kinematics in the disk would require that we describe its size evolution during the collapse through coagulation/fragmentation \citep{lombart_2024}, as well as using a more robust method to treat the dust dynamics in the poorly coupled regime \citep{verrier_2025}. Our results show that in the absence of grain growth, the gas dynamics dominate the distribution of dust and set the baseline metallicity of the disk.

\subsection{Limitations}

Our results must be considered in light of their limitations, whether numerical or due to missing physical processes. While the qualitative picture from our simulations is likely robust, the quantitative details should be interpreted with caution.
\\
The resolution, though sufficient for the collapsing cloud and outer disk, cannot capture the inner disk with adequate fidelity, as too few cells resolve the innermost orbits. The sink particle replacing the sub-AU region further affects the global disk properties, most notably the mass \citep{machida_2014, vorobyov_2019, hennebelle_2020}. We also do not account for high-velocity jets, which can remove substantial amounts of angular momentum \citep{machida_2019}, and the Hall effect, a non-ideal MHD process that can either enhance or suppress magnetic braking \citep{Krasnopolsky_2011, tsukamoto_2015, wurster_2016, marchand_2019}. Furthermore, studies investigating the evolution of the grain distribution throughout the collapse shed doubt on the MRN size distribution \citep{kawasaki_2023, goy_2024, lombart_2024}.
\\
Taken together, these caveats suggest that while our simulations provide valuable insights into disk evolution, they should be viewed as a step toward a more complete physical picture rather than a definitive model.

\section{Conclusion} \label{section:conclusion}

In this paper, we have modeled the self-consistent formation and evolution of protoplanetary disks from the gravitational collapse of $1~\mathrm{M_{\odot}}$ and $3~\mathrm{M_{\odot}}$ cloud cores, using radiative magnetohydrodynamic simulations that include non-ideal magnetic effects (ambipolar diffusion). We follow the evolution of the disks for $\approx 58.5$~kyr and $\approx27$~kyr post sink formation for both simulations. Our analysis show how anisotropic accretion governs both the global properties and internal kinematics of nascent disks, while accounting for eccentric fluid motions to avoid biasing our measurements. Our results can be summarized as follows:
\begin{enumerate}[label=(\roman*)]
    \item Magnetic fields, in tandem with turbulence during the collapse, drive asymmetric accretion via mechanisms like the interchange instability. This process channels material onto the disk primarily through its vertical surfaces, yet the highest mass flux per unit area is from radial infall. This creates a complex accretion geometry that imprints distinct thermodynamic histories on the accreted gas. Gas tracer particles show that gas accreted through polar regions undergoes repeated heating and cooling cycles, while equatorial material follows a more classical adiabatic heating path prior to disk accretion.
    \item Nascent disks exhibit significant and persistent eccentricities, with $e\sim 0.1$. This eccentricity is not transient but is continuously generated and sustained by the anisotropic accretion, which supplies material with a net angular momentum deficit, which sustains eccentricity despite numerical/physical damping.
    \item Vertical accretion drives vigorous turbulent motion throughout the disk, which dominates angular momentum transport. The associated Reynolds stress (with an "effective" $\alpha \approx 0.1$) far exceeds the contribution from gravitational and Maxwell stresses and causes a rapid spreading of the disk with median mass decretion rates $\sim [10^{-5}-10^{-4}]~\mathrm{M_{\odot}~yr^{-1}}$. 
    \item We place our results in the broader context of the observed isotopic dichotomy of Solar System meteorites: The vertical infall distributes material across a broad range of radii during the Class 0 phase, and the Reynold stress it generates enables the outward transport of high-temperature condensates (CAIs, AOAs) without requiring all material to pass through the innermost regions of the disk. Although this can account for the presence of material with non-solar isotopic composition in regions inhabited by CC chondrites, the absence of a sustained inner-disk bias at the early times investigated here suggests that an alternative accretion regime is needed to explain the predominance of NC-like material in the inner disk.
\end{enumerate}
\noindent Our results establish a direct link between magnetically-driven accretion during the collapse and the kinematic properties of protoplanetary disks. The resulting eccentricity and efficient turbulence driving have strong implications for the early stages of planet formation, as they set the orbital conditions for planetesimal growth, as well as for our interpretation of the Solar System's cosmochemical record.

\begin{acknowledgements}
    We thank Matthias González for insightful comments during the writing of this paper. Simulation R1 was carried out by FL, and R2 by BC. This work has received funding from the French Agence Nationale de la Recherche (ANR) through the projects DISKBUILD (ANR-20-CE49-0006), and PROMETHEE (ANR-22-CE31-0020). We gratefully acknowledge the support from the CBPsmn (PSMN, Pôle Scientifique de Modélisation Numérique, \citealp{Quemener13}) at ENS de Lyon for providing computing resources to carry-out and analyze our simulations. EL was supported by ERC No. 864965 (PODCAST) under the EU’s Horizon 2020 programme. A.M. acknowledges the support from ERC grant 101019380 (HolyEarth).
\end{acknowledgements}

\bibliographystyle{aa}
\bibliography{biblio}

\begin{appendix}
\section{Disk definition} \label{appendix:diskdef}

\begin{figure}[h]
\centering
\includegraphics[scale=.34]{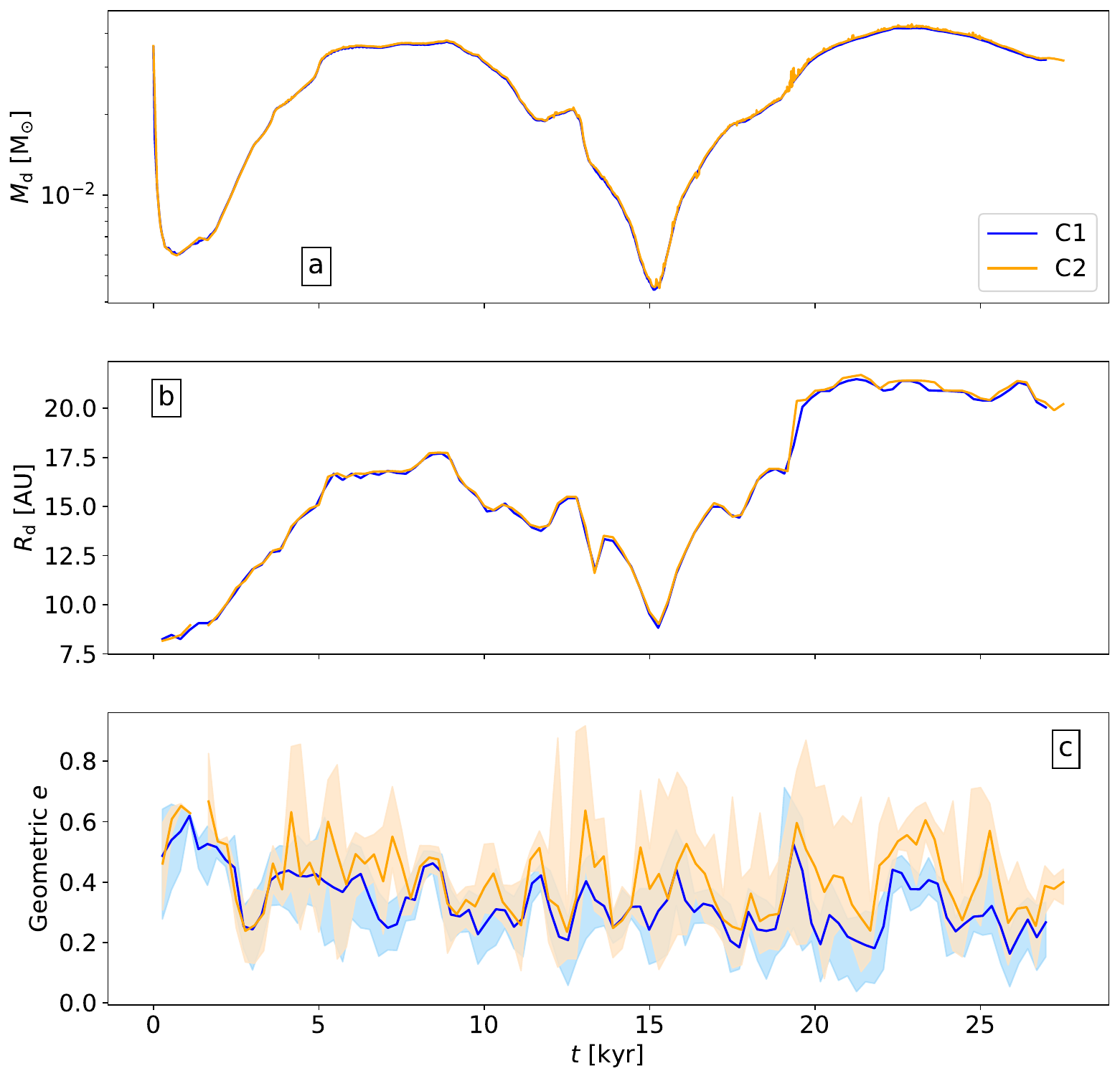}
\caption{Comparison of criteria C1 (blue) and C2 (orange) in run R2 (see text). Shown here are disk mass (a), 90\% mass radius (b), and apparent eccentricity (c) as a function of time.}
\label{fig:diskdefcomp}
\end{figure}

Following \cite{commercon_2024}, we define the disk using a local angular momentum criterion: all cells with angular momentum exceeding $5\times10^{-3}\ |\vec{L}|_{\mathrm{max}}$ (where $|\vec{L}|_{\mathrm{max}}$ is the maximum local angular momentum in the simulation domain) are identified as disk material. An alternative criterion from \cite{lee_2021} selects cells above the minimum disk density. We compare these two definitions (respectively C1 and C2) in Fig.~\ref{fig:diskdefcomp}, which shows that while disk mass and radius agree well, C2 yields consistently higher apparent eccentricities as it better captures the outermost orbit. For simplicity, we adopt C1.

\section{Counter-rotating streamer} \label{appendix:shrink}

\begin{figure}[h]
\centering
\includegraphics[scale=.5]{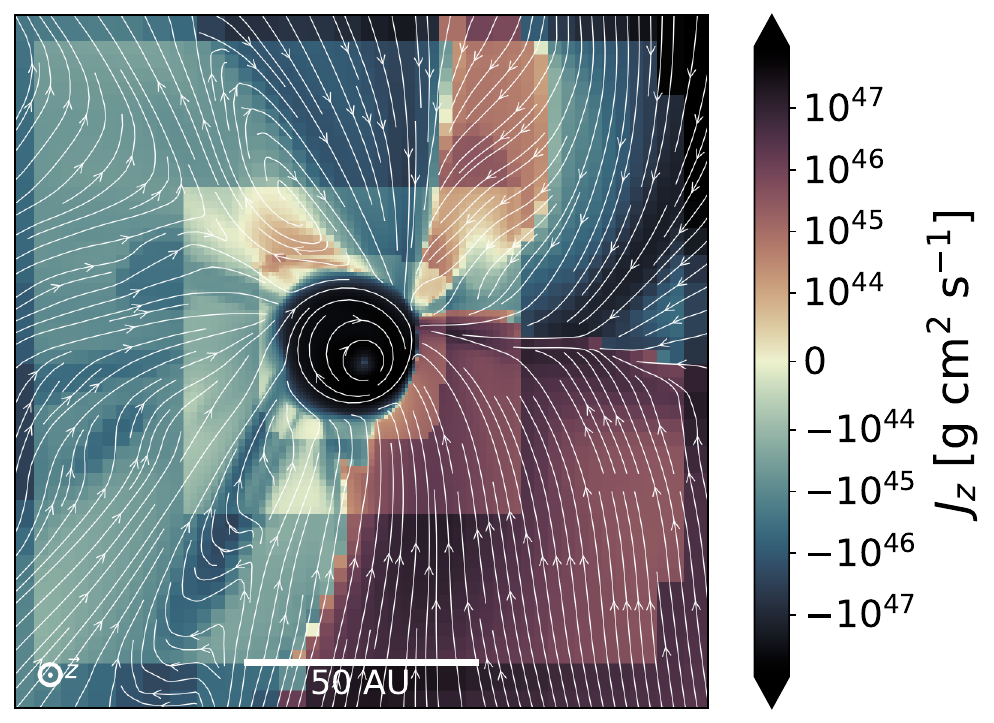} 
\caption{Slice in the $z$-direction of the $z$-component of the angular momentum in run R2 at $t \sim 13$ kyr, showing a counter-rotating streamer delivering material to the disk. The white curves are velocity vector field streamlines.}
\label{fig:streamerJ}
\end{figure}

In Fig. \ref{fig:globalprops}b, the disk radius in run R2 (red curve) shows a dip around $t \approx 15$ kyr, caused by accretion from a counter-rotating streamer. Figure \ref{fig:streamerJ} displays the $z$-component of the angular momentum,
\begin{equation}
\vec{J} = \vec{r} \times m \vec{v},
\end{equation}
(where $m$ is the mass within a cell) at $t \sim 13$ kyr, highlighting the streamer with opposite angular momentum. The influx of such material forces the disk to contract.

\section{Edge-on view of accretion} \label{appendix:EOaccr}

\begin{figure}[h]
\centering
\includegraphics[scale=.5]{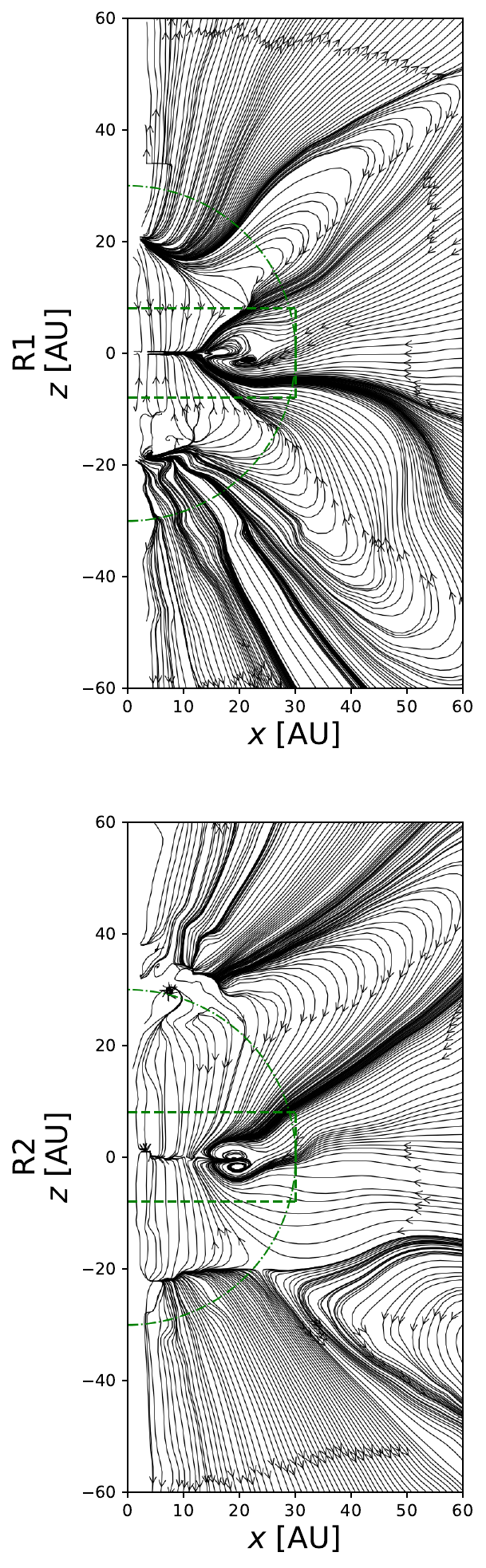} 
\caption{Azimuthally averaged velocity vector field streamlines of runs R1 (top) and R2 (bottom) at our final simulation snapshots. The green dashed and dash-dotted lines in each plot displays the extents of the fixed surfaces used to quantify accretion in Sections~\ref{sec:dirmf} and~\ref{sec:cylmf}.}
\label{fig:eoaccr}
\end{figure}

We have shown in Sections \ref{sec:dirmf} and \ref{sec:cylmf} the origin of the accreted material that reaches the disk. This appendix further illustrates those results by presenting velocity vector field streamlines.  
\\
In Figure \ref{fig:eoaccr} we show the azimuthally averaged velocity vector field streamlines at the final simulation snapshots for runs R1 (top) and R2 (bottom). These highlight the outflow, the infall beneath the outflow, and the magnetically regulated infall. The extents of the spherical and cylindrical surfaces used in Sections~\ref{sec:dirmf} and~\ref{sec:cylmf} are indicated by dash-dotted and dashed green lines, respectively.

\section{Gravitational stability of the disk} \label{appendix:toomre}

\begin{figure}[h]
\centering
\includegraphics[scale=.5]{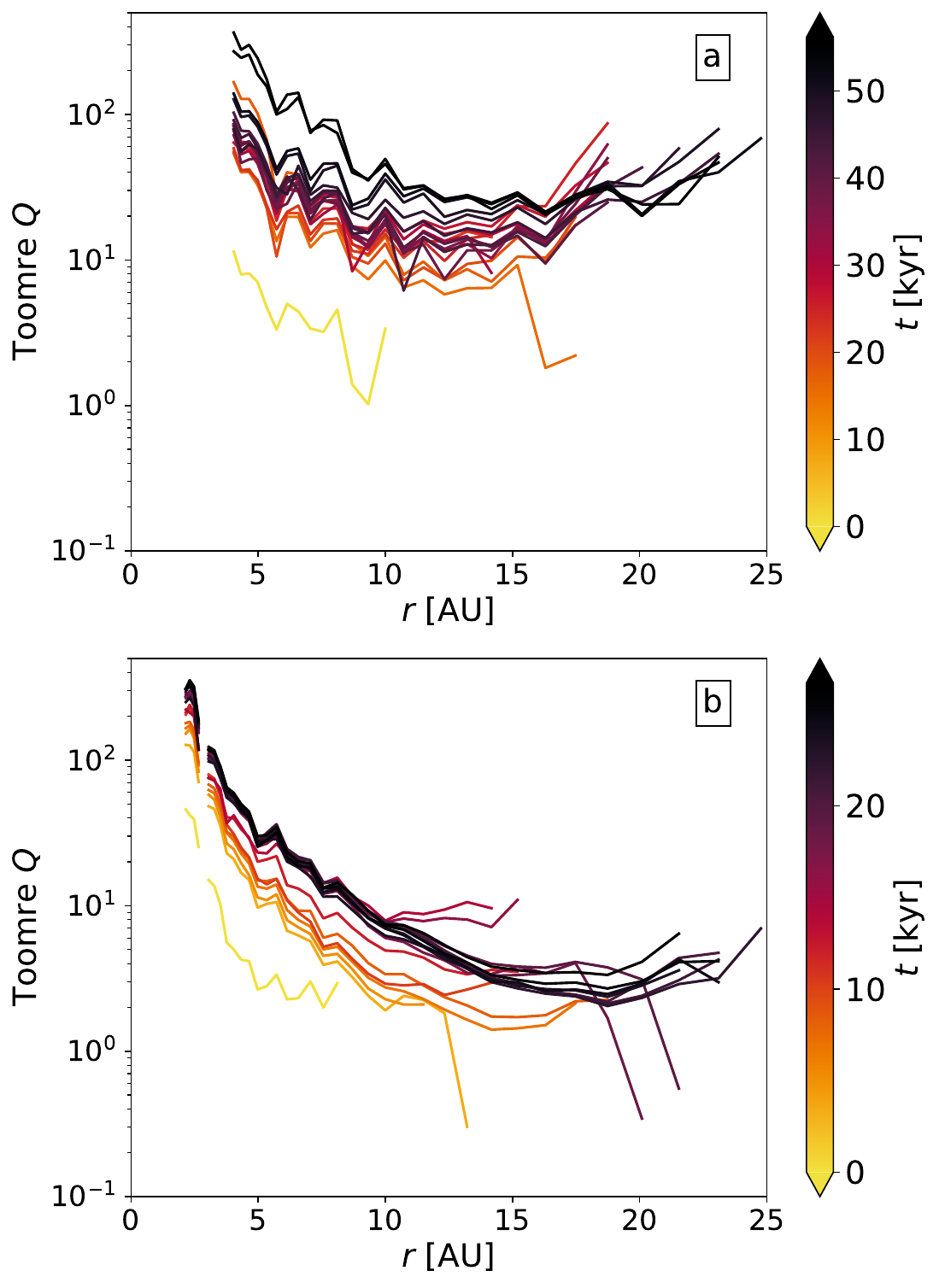} 
\caption{Averaged values of the real part of Toomre $Q$ for runs R1 (a) and R2 (b) as a function of radius, where each curve corresponds to a different time, with $t=0$ denoting the moment of sink formation. Only cells belonging to the disk were used when computing these curves.}
\label{fig:toomre}
\end{figure}

We analyze the gravitational stability of our disks using the classical Toomre parameter \citep{Toomre_1964}, which measures the ratio of thermal and centrifugal support to self‐gravity. A value of $Q>1$ indicates that the disk is stable against axisymmetric gravitational perturbations.
\\
Figure \ref{fig:toomre} shows the radially averaged Toomre profiles for both runs. At all times in both simulations, we find $Q\gg1$, indicating strong stability throughout the simulation's duration, particularly at later times. The more massive disk in Run 2 (R2) exhibits systematically lower values of $Q$ than the disk in R1, yet remains well above the instability threshold.

\section{Density and temperature profiles} \label{appendix:avprofilesrhoT}

\begin{figure*}[h]
\centering
\includegraphics[scale=.5]{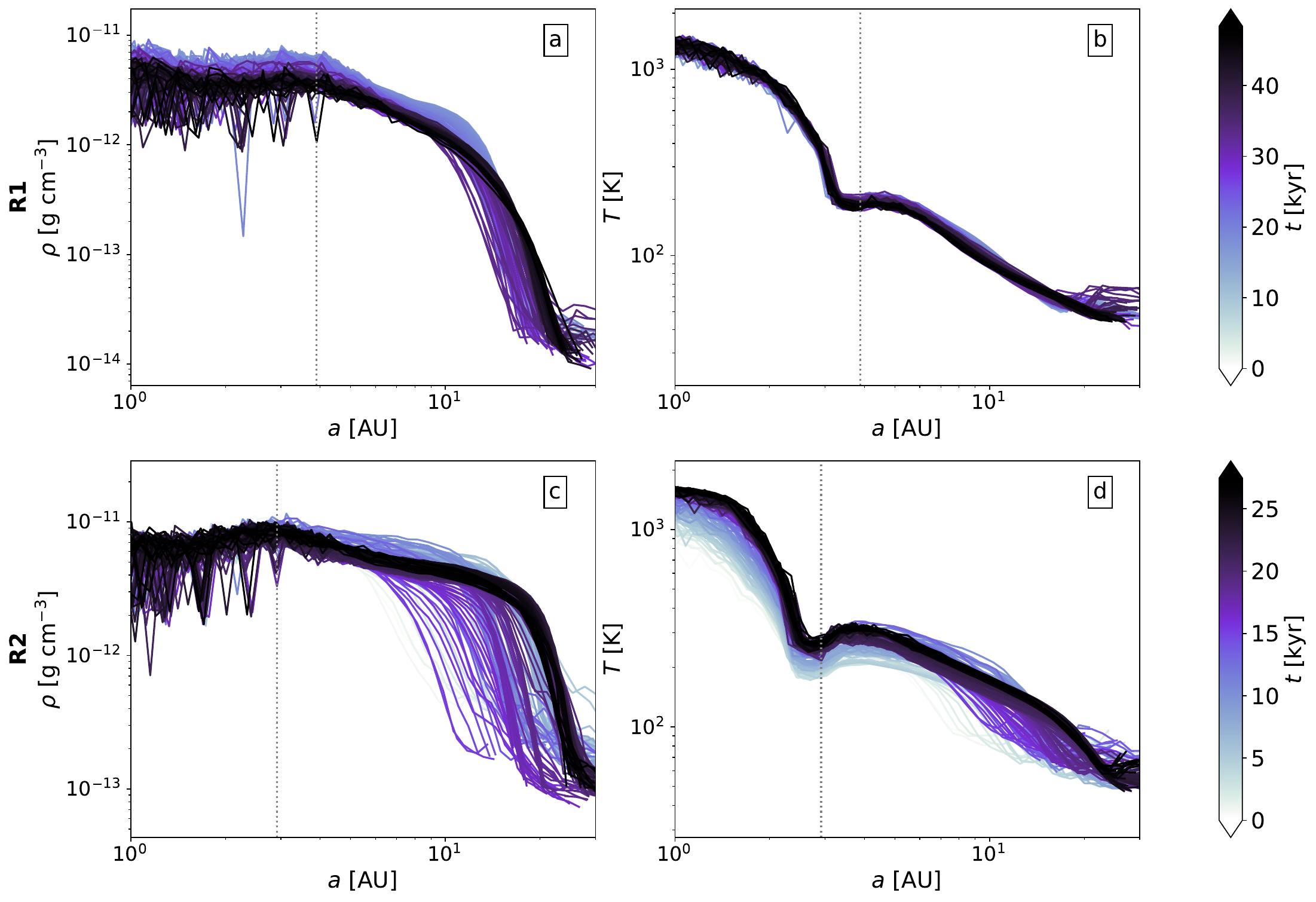} 
\caption{Mass-weighted averages of the disk density (a, c) and temperature (b, d) for runs R1 (a, b) and R2 (c, d), where each curve denotes a different time, with $t=0$ corresponding to the moment of sink formation. The gray vertical dotted lines correspond to the accretion radius of the sink particle.}
\label{fig:avrhoT}
\end{figure*}

The thermal history of the gas in the midplane has consequences on the mineralogy of the first condensates \citep{charnoz_2024}. Figure \ref{fig:avrhoT} shows the mass-weighted density and temperature profiles for the disk midplanes in runs R1 and R2. For R1, both density and temperature decrease gradually over time. In contrast, R2 exhibits a more dynamic and chaotic history: episodes of disk accretion cause temporary increases in density and temperature, which are followed by cooling during quieter periods. Averages are computed using only cells belonging to the disk.

\end{appendix}
\end{document}